%% file: main.tex
\documentclass[letterpaper,twocolumn,10pt]{article}
\usepackage{usenix-2020-09-old}

\usepackage[english]{babel}
\usepackage{graphicx}
\usepackage{subcaption}
\usepackage{booktabs}
\usepackage{longtable}
\usepackage{listings}
\usepackage{amsmath}
\usepackage{comment}
\usepackage{tcolorbox}
\usepackage{array}
\usepackage[table]{xcolor}   
\usepackage{titlesec}       
\usepackage{soul}
\usepackage[textsize=small]{todonotes}
\usepackage{tabularray}     
\usepackage[normalem]{ulem}
\usepackage{hyperref}
\newcolumntype{P}[1]{>{\centering\arraybackslash}p{#1}}

\titlespacing*{\subsection}{2pt}{8pt}{8pt}
\titlespacing*{\subsubsection}{2pt}{6pt}{6pt}

\newcommand{\squishlist}
{
    \begin{list}{$\bullet$}
    {
        \setlength{\itemsep}{0pt}      \setlength{\parsep}{2pt}
        \setlength{\topsep}{2pt}       \setlength{\partopsep}{0pt}
        \setlength{\leftmargin}{1em} \setlength{\labelwidth}{0.5em}
        \setlength{\labelsep}{0.5em}
    }
}
\newcommand{\squishend}
{
    \end{list}
}
\newcommand{\name}{SILC}

\input{macros}

\makeatletter
\g@addto@macro\@maketitle{\vspace{0.35in}}
\makeatother

\begin{document}

\title{\Large \name: Lookahead Caching for Short-form Video Delivery Systems}

\author{
\begin{tabular}{c}
Maleeha Masood$^{1}$ \quad Shreya Kannan$^{1}$ \quad Om Chabra$^{2}$ \quad
Deepak Vasisht$^{1}$ \quad Indranil Gupta$^{1}$\\
{$^{1}$University of Illinois Urbana-Champaign \quad
$^{2}$Massachusetts Institute of Technology}
\end{tabular}
} 

\maketitle

\begin{abstract}
\input{abstract}
\end{abstract}

\input{intro-3}
\input{primer}
\input{overview}
\input{distribution}
\input{design}
\input{user-study}

\input{eval-2}
\input{related}
\input{conclusion}

\bibliographystyle{acm}
\bibliography{reference}

\input{appendix}
\end{document}

%% file: macros.tex
\newtoggle{our_comments}
\newtoggle{our_contents}
\toggletrue{our_comments}
\toggletrue{our_contents}

\newcommand{\para}[1]{\vspace{4pt}\noindent\textbf{#1}}

\newcommand{\rw}[1]{
\iftoggle{our_comments}{%
}{}}

%% file: abstract.tex
Short video platforms like TikTok, Instagram Reels, and YouTube Shorts have gained immense popularity in the last few years and are responsible for a large and growing fraction of Internet traffic. We identify two unique opportunities for improving short video delivery using their existing interactions with content delivery networks (CDNs). First, short videos use a push-based recommendation system, where the user is presented a sequence of videos recommended by the algorithm rather than 
user explicitly picking content to watch (e.g., in YouTube). Such push-based short video systems offer a unique opportunity for system design by providing visibility into upcoming requests. 
Second, the popularity of these videos follows a highly skewed Pareto distribution, leading to geographical and temporal overlap amongst videos being served.
We leverage these opportunities to build \name---a {\it lookahead-aware} caching system, aimed at (i) reducing CDN cache miss rates,  as well as (ii) reducing midgress bandwidth between the CDN and the origin server. 
Our evaluation of \name\ uses traces that we collect from real users, through (i) an in-person user study, and (ii) a data donation program involving 100 TikTok users across the world.
Using a combination of these traces, we simulate traffic from 10,000 simultaneous users. Our evaluation shows that,  compared to 10 state-of-the-art heuristic and learning-based cache eviction policies,  \name\ 
reduces a CDN's midgress costs by 11.1\% to 111\%. 

%% file: intro-3.tex
\section{Introduction}\label{sec:intro}

\begin{figure}
    \centering
    \includegraphics[width=0.18\textwidth]{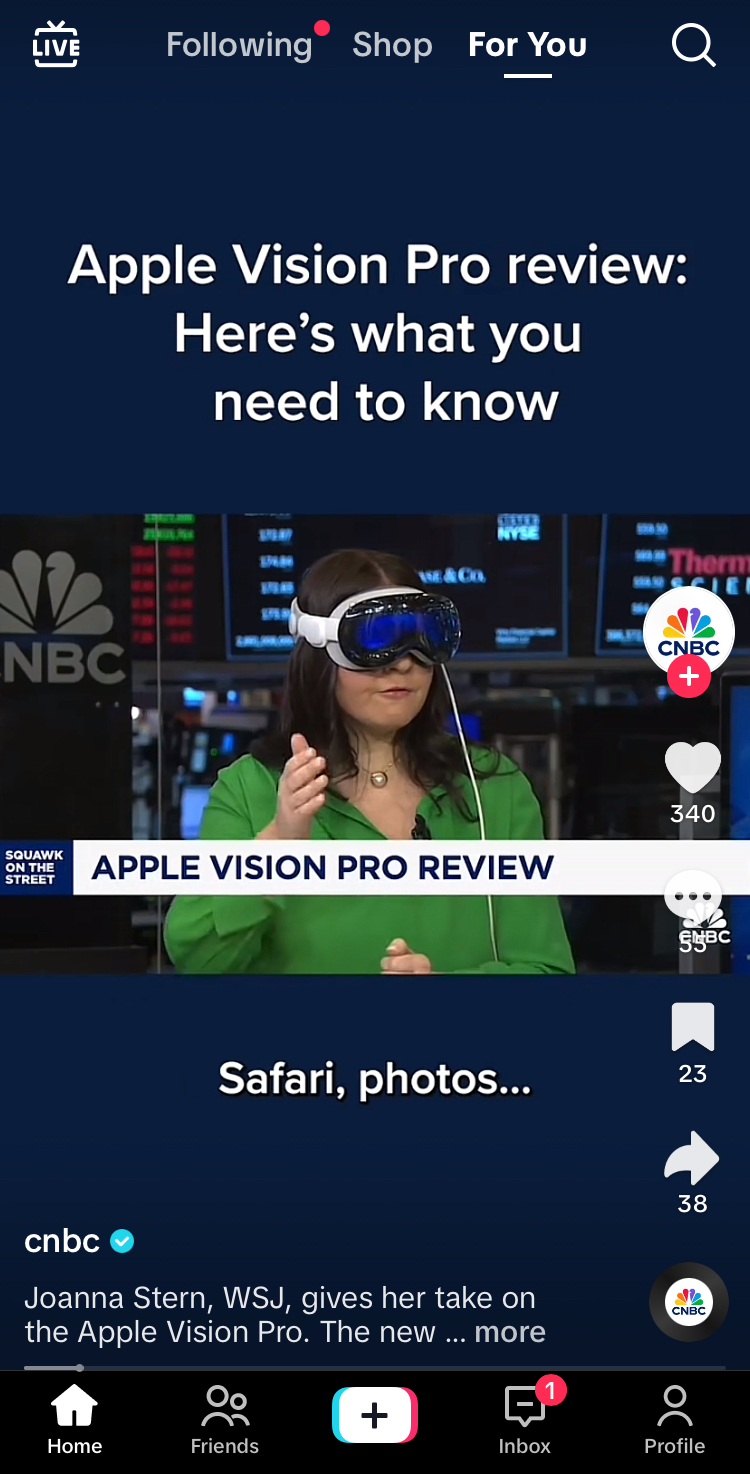}\quad\quad\quad
    \includegraphics[width=0.18\textwidth]{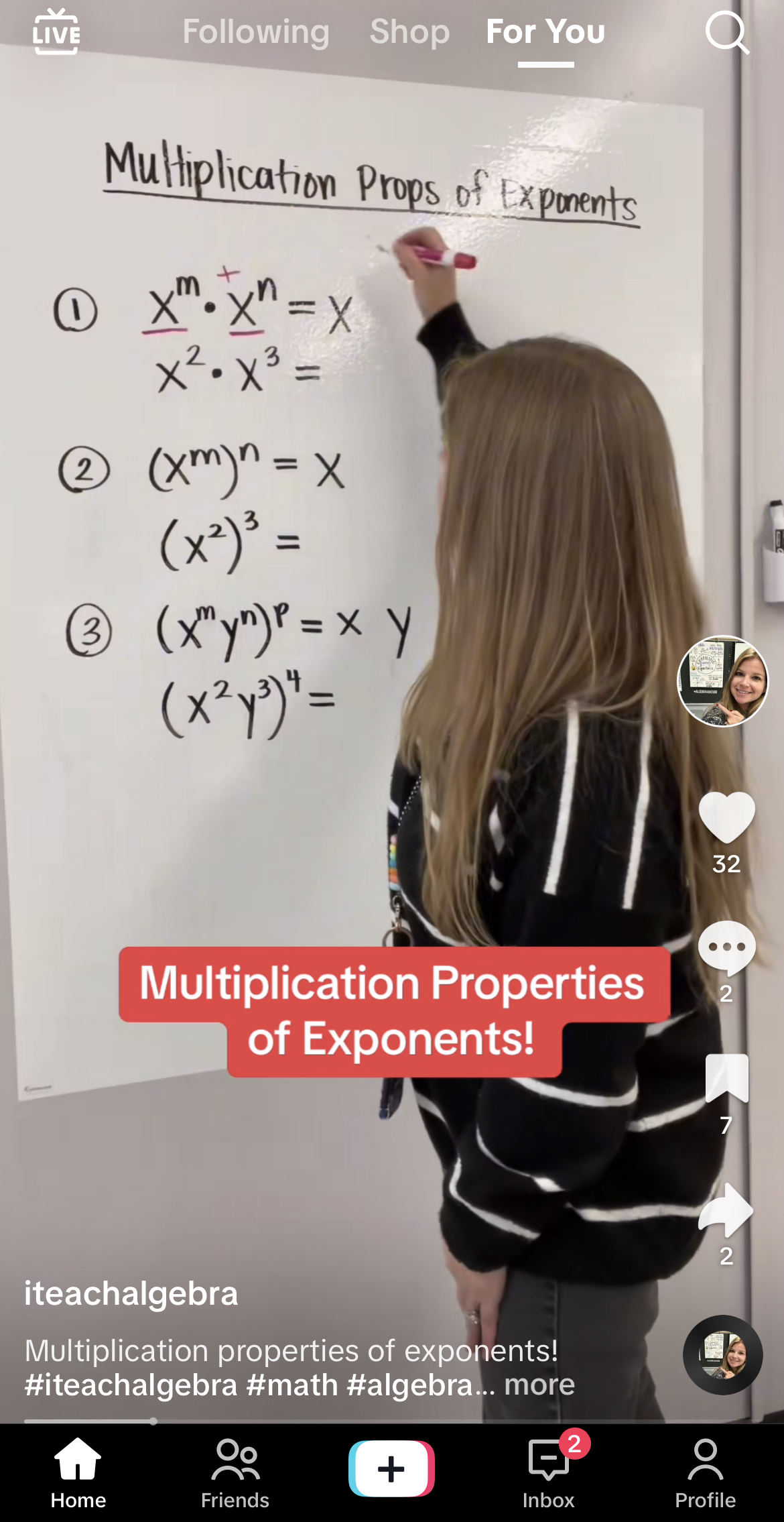}\vspace{-0.1in}
    \caption{\centering \textbf{TikTok's FYP (For You Page).} \textit{Users are shown a sequence of videos tailored to their personal interests as learned by the recommendation algorithm.}}
    \label{fig:tiktok-fyp}
\end{figure}

The last few years have seen an explosion of short video delivery systems like TikTok, YouTube Shorts, Instagram Reels, and Douyin. TikTok has 1 billion~\cite{tiktokstat} monthly active users,  while Youtube Shorts and Instagram have 2 billion~\cite{ytstat}, and 1.8 billion~\cite{instastat} respectively. In addition to entertainment, these short video systems are increasingly becoming platforms for microlearning (e.g., quizzes, school tutorials, how-to videos), citizen reporting, advertising, user-generated content, testimonials, and sports \cite{khlaif2021using, wu2021study}. A  growing body of research has begun  investigating how to optimize short-video streaming, with a particular focus on problems such as adaptive bitrate algorithms, video quality management, and network bandwidth reduction \cite{nctmwww24, winwin, tladder, personalizedadaptive}.

In this paper, we focus on the problem of caching in CDNs (content delivery networks) for short video platforms. CDNs are a critical part of the delivery stack---TikTok, for example, relies on more than 1000 CDN nodes (e.g., Fastly, Akamai, and its own network) to serve content cached geographically near a user (typically via HTTP) \cite{tiktokcdn}. However, inefficient caching mechanisms at CDNs result in  high costs for  the CDN operator, and hurt user experience. First, a cache miss forces the video to be fetched from a distant origin server, incurring significantly higher latency than a cache hit and directly degrading a user's playback quality.

Second, CDNs charge platform providers such as  TikTok/ByteDance,  YouTube/Alphabet, Instagram/Meta for {\it egress} traffic, the traffic sent from the CDN to end users. In contract, CDNs themselves bear the cost of {\it midgress traffic}, which arises whenever an uncached video must be fetched from the origin server to the CDN. 
Past work \cite{sundarrajan2020midgress} has estimated that midgress costs for CDNs surpass \$60 million per year---thus techniques that reduces midgress by even a fraction of a percentage point, can signficantly increase  profit margins. 
While CDN caching for long-form video (e.g., YouTube) has been studied \cite{krishnappa2015optimizing, ghaznavi2021content}, 
relatively little research exists in optimizing CDN design for the emerging class of short video delivery systems.

We observe two unique characteristics of short video systems (compared to previous long video systems) that offer new opportunities for designing CDNs: 

\para{1. Short Video Systems Utilize {\it \textbf{Push-based}} Recommendations: }  
In traditional long-form streaming like YouTube and Netflix, users actively search for and select videos to watch (even from among the recommendations on the home page). In other words, users {\it pull} content by telling the system which videos to fetch. However, in short video systems, the {\it recommendation algorithm} decides what the users watch to maximize user engagement. Hence, short video systems are {\it push-based recommendation} systems. The algorithm presents each user with a list of videos that the user views sequentially, in the order presented by the algorithm---each video can be viewed completely, or skipped after watching partially. The user may also choose to fast forward/rewind the video, or (less common) go back to the previous video. 
These interactions are then used as feedback by  the recommendation algorithm.  

For example, on TikTok, users primarily engage with 
content through the “For You Page (FYP)”, swiping through a continuous stream of videos chosen by the recommendation algorithm (see Fig.~\ref{fig:tiktok-fyp} for a sample user-view of the FYP). Unlike traditional platforms where users explicitly select what to watch, this push-based model gives the system advance knowledge of which videos will appear next. This lookahead creates a {\it new system-design opportunity}: while today’s CDNs are {\it reactive}—relying only on past requests to predict future demand—short-video platforms expose near-future requests directly. Our work aims to leverage this foresight to make CDN caching {\it proactive}, allowing the system to extract and ameliorate system-level efficiencies.

\begin{figure}
    \centering
    \includegraphics[width=0.45\textwidth]{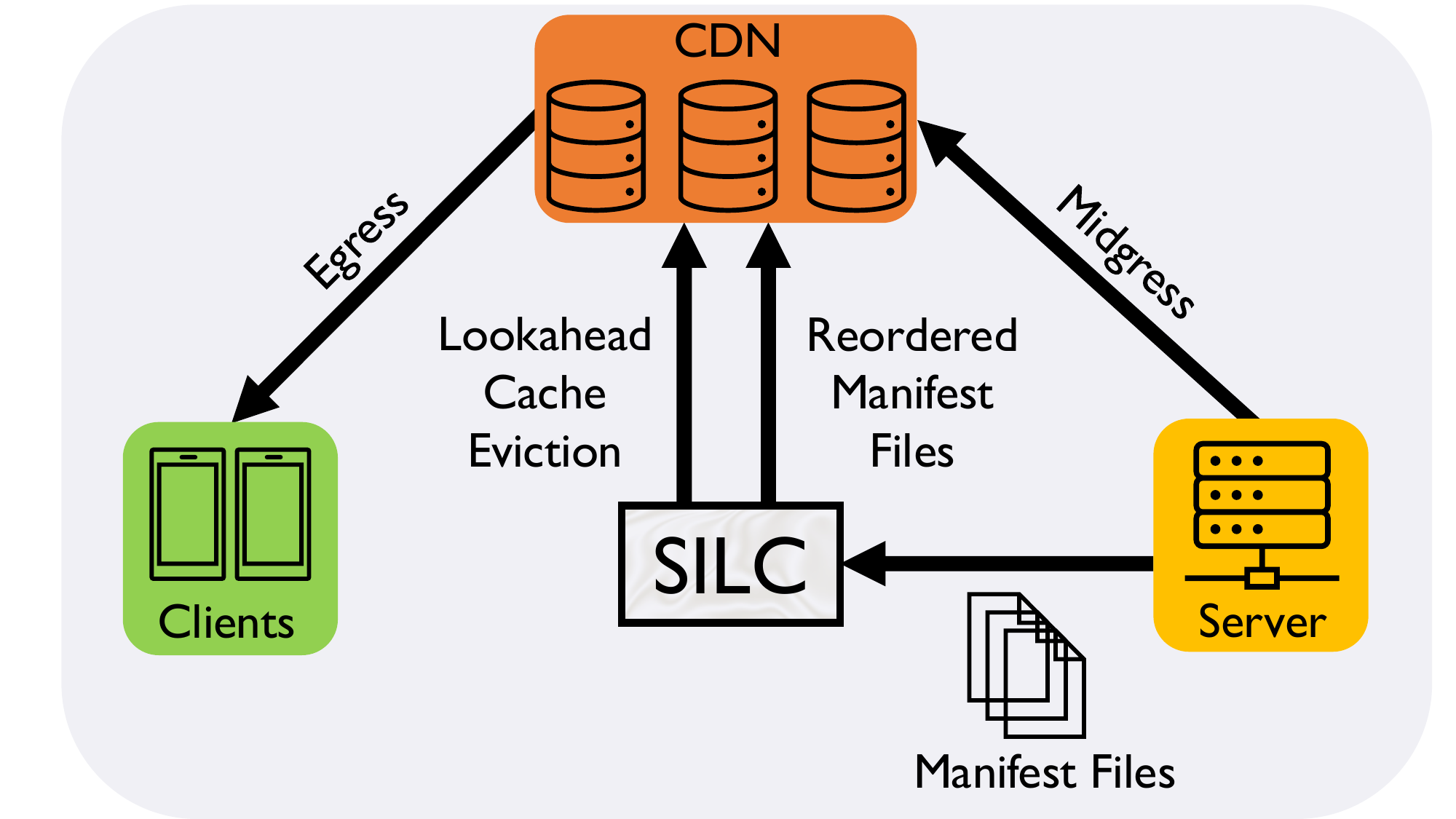}
    \caption{\centering \textbf{System Overview.} \textit{\name\ reduces midgress traffic and improves CDN hit rates through new {\it lookahead eviction} and {\it online reordering}  policies, while preserving user engagement. 
    }}
    \label{fig:overview}
\end{figure}

\para{2. Popularity Distributions in Short Video Systems are Highly Skewed: } 
We organize a data donation program to collect a dataset of 2.65M videos recommended by TikTok to real users between Nov 2023 - Sept 2024. 
Using this dataset, we characterize the popularity distribution of short videos. Our analysis reveals that TikTok video popularity follows a Pareto distribution: approximately 20\% of the videos account for 80\% of the views. We observe this distribution both at the population level and at the per-user level~\footnote{A similar skew has been reported previously in other user-generated content platforms such as YouTube \cite{cha2007tube}.}. Notably, this Pareto distribution is more skewed than the Zipf distributions observed in web traffic \cite{breslau1999web}, data center traffic \cite{zipfcontainer,jiang2020characterizing}, and peer-to-peer file sharing systems \cite{gummadi2003measurement}. In this more extreme skew, a disproportionately small fraction of videos accounts for the majority of bandwidth consumption.

This skewness indicates high potential of caching at the CDN. Since \textit{viral} videos are fetched by many users, caching them would increase CDN hit rate and reduce CDN midgress consumption. One may wonder---why not just cache the top viral videos at the CDN and serve a subset of them to all clients served by the CDN? While we present experimental comparisons against such a strategy, we note here that the most popular videos ``churn'' very quickly in short video systems, as well as because popularity is more ``partitioned'' across users (vs. say webpages)---basically, most short video users see only a fraction of the most popular videos (e.g., geographically correlated with where they currently are located), rather than  all popular videos (like with webpages).

Motivated by these unique characteristics, we design \name{} ({\it \textbf{S}hort V\textbf{i}deo \textbf{L}ookahead \textbf{C}aching}), a new system designed to exploit both the lookahead capability and the high skew in popularity distributions, with the twin aims of (1) reducing CDN misses and CDN midgress consumption (and thereby lowering the operational expenses of CDN operators, {as well as (2) lowering user-perceived miss rates})---all while preserving user engagement. 

The key components in \name\ (Fig.~\ref{fig:overview}) are:

\para{(a) Lookahead-based Cache Eviction:} In short video  systems, the 
content server (e.g., TikTok/YouTube server) currently provides both the the CDN and client a small ordered list of the ``next'' videos 
shown to the user. In TikTok, for instance, this list is called a {\it manifest file} and our measurements showed that a single manifest file contains roughly 30 videos (YouTube and Instagram's equivalent have about 10-15 videos).  
The manifest file contains video metadata (with CDN links), but not videos themselves. 

The CDN can leverage such \textit{short-term future} information to better identify which videos are best to be evicted from the CDN cache. This is unlike existing cache eviction policies like least recently used (LRU), least frequently used (LFU), or other eviction policies which rely on past metrics to predict future popularity. Specifically, we propose \textit{Least Lookahead Frequency (LLF)}, a cache eviction policy that evicts videos that have the least number of near-future views. Since a CDN has manifest files for multiple users, cache eviction decisions using LLF are more accurate because they are based on accurate near-future needs across multiple users.

\para{(b) Reordering to Maximize Overlap:} Given the observed Pareto distribution in video popularity, we expect that many users served by a CDN will have large overlaps in the videos they watch. In fact, a CDN node's geographic locality further amplifies this effect (e.g., popular videos within a city or region). \name\ combines its LLF cache eviction policy with a new {\it online reordering scheme} to further leverage the temporal overlap between users. Specifically, in \name, the CDN has access to manifest files of multiple users being served by it. Therefore, it can reorder the videos in the manifest files to maximize the temporal overlap such that requests for a video are clustered in time. This improves cache utilization by reducing the time in cache for each video. In practice, this is challenging because the manifest files only contain the order of videos and not the exact time a video will be played (users can skip or replay videos). \name\ optimizes for temporal overlap despite this uncertainty by combining the knowledge of CDN's current cache state with the information in manifest files.  

While \name\ introduces cost savings for CDNs (and potentially, content providers), a key concern arises regarding the impact of reordering videos on the user experience. 
 
To rigorously measure the engagement impact of \name's reordering component, we conduct a user study that performs a controlled within-subjects \cite{withinsub} A/B test (also known as Comparative Usability Testing~\cite{comparativetesting}). Our study consisted of 44 participants, each viewing both their original (unmodified) feed and a reordered feed and then rating their engagement on a Likert Scale (1-7). Our results showed that the two feeds were statistically indistinguishable, and participants could not reliably discriminate the reordered condition, indicating that reordering does not substantially degrade user engagement. 

\label{sec:principleduserstudies}

\name\  makes the following contributions:
\squishlist
    \item We present the first analysis of short video popularity (focusing on TikTok), showing it 
    follows a skewed Pareto distribution. 
    \item We design a new \textit{Least Lookahead Frequency (LLF)} cache eviction policy for short videos, aimed at reducing CDN midgress. 
    \item We present a new {\it online} video reordering algorithm which maximizes video overlap across users {(without changing the {\it set} of per-user recommendations)} and thus reduce CDN midgress. 
    We verify that this reordering does not impact user's engagement 
    using the methodology of {\it within-subjects A/B testing} (also called Comparative Usability Testing). 
    \item We leverage a novel design and experimental methodology that includes user studies to motivate our key research questions, data donations to collect a large dataset, and an experimental system evaluation driven by traces from real participant behavior. 
    \item {Our experimental results utilize 2.65M videos collected from 100 TikTok users. They show that \name\  reduces a CDN's midgress costs by 11.1\% to 111\% compared to 10 state-of-the-art heuristic and learning-based cache eviction policies. Compared to a practical implementation of Belady's MIN algorithm \cite{beladymin}---known to be  optimal when {\it all} future requests are known---when both operate only on the limited lookahead available from manifest files, \name\ still reduces midgress traffic by at least 12.3\%.}
\squishend

%% file: primer.tex
\vspace{-5pt}\section{Primer and Background}

We describe how the TikTok app works and the role CDNs play in delivering short video content. 

\para{TikTok -- }When a user opens the app, they see the “For You Page” (FYP) (user view of the app shown in Fig.~\ref{fig:tiktok-fyp}). The FYP page presents the first short video, and the user can choose to either watch it completely (after which it loops), or “swipe up”. Swiping up presents the next recommended video. The user could also swipe down to go back to the previously watched video (this is less common). Users cannot skip videos entirely and see at least a (small) part of the video as they swipe if they do not want to watch a video. The user can engage with each video by liking it, leaving a comment, sharing it, visiting the creator’s profile, or other actions (bookmark, download, etc.). In our experience, the most common user activity is watching the stream of recommended FYP videos. The push-based recommendation algorithm that generates the FYP video sequence is at the heart of why such short video platforms are popular (these recommendation algorithms are proprietary).

Short videos on TikTok are User Generated Content (UGC) uploaded to the platform by users. Any user on TikTok can upload a video - currently, most users can upload videos under 10 minutes (recently increased to 60 minutes). In our dataset, we observe that 75\% of videos are less than a minute long. 
Once uploaded, TikTok's recommendation algorithm  decides who to recommend videos to. Authors may choose to sponsor their content for more views by paying TikTok \cite{Liang2021} but most popular videos are typically not sponsored.

\para{CDNs -- } CDNs serve a large fraction of Internet content today \cite{stocker2017growing}. CDNs are distributed globally and serve as local caches for video providers such as Netflix, YouTube, and TikTok \cite{wang2018comparing}. Clients request objects from CDNs that maintain a local cache of popular content. If the CDN has a copy of the object (\textit{CDN hit}), it immediately delivers it to the client. However, if the CDN does not have a copy (\textit{CDN miss}), it fetches the object from the origin server and delivers it to the client. A CDN miss incurs additional delay because of the additional connection from the CDN to the server. In addition, the CDN incurs additional cost in terms of midgress bandwidth on a miss, i.e., consuming bandwidth to get content from the server to the CDN to respond to client requests. 

%% file: overview.tex
\section{\name\ Design}

\name\ redesigns how CDNs serve short videos to users. Given the user manifest file (from the content provider/client), \name\  obtains and leverages near-future knowledge. We seek to achieve the following goals:
\squishlist
    \item \textit{Lower Midgress: } 
    CDN midgress is related to the miss rate, but also captures the bandwidth impact of misses (e.g., missing bigger videos causes more impact to midgress). Midgress costs are incurred by the CDN but not paid by the content provider (e.g., TikTok), and need to be minimized.
    \item \textit{Higher Hit Rate at CDN}:  CDN misses lower user quality of experience and platform engagement by increasing delays. We aim to reduce CDN misses.
    \item \textit{Recomendation-agnosticity: }We seek to remain faithful to the existing recommendation algorithm. Specifically, we do not aim to change the {\it set of} videos recommended by the algorithm, treating the latter as a black box.
    \item \textit{Transparency to Users: }\name\ modifies the delivery of short videos but is designed not to impact the engagement of users with TikTok's recommended videos. 
\squishend

\S\ref{sec:analysis} begins by analyzing TikTok network traffic and popularity characteristics of TikTok videos. Based on these observations, we design \name's new cache-eviction policy (\S\ref{sec:llf}) and online manifest file reordering scheme (\S\ref{sec:reordering}). We validate our design choices and impact on user engagement via user studies in \S\ref{sec:userstudy}. Finally, we evaluate \name\ in \S\ref{sec:results}. 

%% file: distribution.tex
\subsection{Unique Network Traffic Patterns in Short Videos}\label{sec:analysis}

In push-based recommendation approach that short video systems use, the videos seen by users are driven by a backend recommendation algorithm that fine-tunes itself to each user's interests. We describe the mechanism that TikTok uses to convey these recommendations first, and then analyze it by using TikTok's network traffic and a large corpus of TikTok videos that we collected.

\lstset{ captionpos=b,
frame=lines, showspaces=false, showtabs=false, breaklines=true,
showstringspaces=false,
escapeinside={(*@}{@*)}, commentstyle=\color{greencomments},
morekeywords={partial, var, value, get, set},
basicstyle=\ttfamily\scriptsize, }
\begin{figure}[t]
\begin{lstlisting}[language=python]
{
   "extra": {"fatal_item_ids": [], "logid": "2023050719530375D121B900725B73BC5D", "now": 1683489185000},
    "hasMore": true,
    "itemList": [
        {..., "forFriend": false, "id": "7382032470033698080", ...},
        {..., "forFriend": false, "id": "7368220166003510570", ...},
        ...
    ],
    "log_pb": {"impr_id": "2023050719530375D121B900725B73BC5D"},
}
\end{lstlisting}\vspace{-0.1in}
\caption{\centering \textbf{Example Manifest File.} \textit{A manifest file contains around 30 videos in the itemList and decides the sequence in which videos appear on a user's FYP. }}\vspace{-15pt}
  \label{fig:manifestfile}
\end{figure}
\subsubsection{Manifest Files} 
We intercepted HTTPS communication for TikTok's web application using a proxy tool called mitmproxy \cite{AdamMitm}. Based on our analysis, we detail below the flow of control (recommendations) and data (videos) between user client device, CDN, and the TikTok content server. 
 First, the TikTok client initiates a connection with the content (TikTok) server. The content server sends a {\it manifest file} for that client to the CDN, and this manifest file is sent forward to the client.

This manifest file contains an {\it ordered} list of recommended videos and their URLs, {as recommended by the TikTok's algorithm for that individual user}. Fig.~\ref{fig:manifestfile} shows a snippet of a manifest file from TikTok.  The client app (at user) then {\it sequentially} downloads these videos---in the order they appear in the manifest file---from the CDN, and displays them to the user. User viewing and file downloading occur in parallel (e.g., TikTok fetches the next four videos from the manifest file) \cite{li2023dashlet}.

When a client is done viewing most of the videos in a manifest file (with 5-10 remaining to be seen), the client requests a new manifest file in the background. Then the above process repeats. 
TikTok's use of a manifest file has also been noted previously \cite{li2023dashlet}, and {our findings above reflect the latest version of TikTok as of September 2025.   

While the use of recommendation lists is common in long-form video streaming systems like YouTube, a user can choose one (or a few video) from the set of recommendations. In contrast, a user sequentially watches (at least the first part) of each recommened video in a push-based video recommendation systems used in short video platforms.

\para{Other Platform's Manifest Files:}
Other short video applications, like Instagram Reels and YouTube Shorts also utilize manifest files to manage video recommendations - we use mitmproxy to intercept these files. Instagram's manifest file typically includes a list of 10–15 videos. As users approach the end of their current list of reels, a new manifest file is generated. These files contain metadata about each reel, such as its popularity, duration, and author details. Similarly, YouTube Shorts employs manifest files, but with a simpler structure compared to Instagram. YouTube's manifest files primarily contain information about the next 15 shorts, such as their URLs or video IDs, without as much detailed metadata. The manifest files for both Instagram and YouTube are very similar in structure to that of TikTok. Thus, the schemes we apply to the TikTok manifest files can also be applied to Instagram Reels and YouTube Shorts~\cite{instacdn,ytcdn}.

\para{Implication:} The existence of the manifest file and the sequential display of the videos in the file to the user in a push-based recommendation system, together,  enable a unique opportunity for system design. We can now \textit{lookahead} into the future, i.e., the manifest file tells the CDN the next list of videos that will be served to the user's app. The CDN can use this information to learn about near-future requests and improve its cache hit rate and midgress by making informed eviction decisions. 

\vspace{3pt}\noindent{\it Observation 1: Manifest files allow CDNs to learn about future requests.}\vspace{3pt}

\subsubsection{Pareto Popularity Distribution}

\lstset{ captionpos=b,
frame=lines, showspaces=false, showtabs=false, breaklines=true,
showstringspaces=false,
escapeinside={(*@}{@*)}, commentstyle=\color{greencomments},
morekeywords={partial, var, value, get, set},
basicstyle=\ttfamily\scriptsize, }
\begin{figure}[t]
\begin{lstlisting}[language=python]
{
    "id": 7492338560951913733, 
    "desc": ..., 
    "createTime": ..., 
    "scheduleTime": ..., 
    "video": ..., 
    "author": {}, 
    "music": ..., 
    "stats": {"diggCount":277800, "shareCount":11900, "commentCount":77300, "playCount":21200000, "collectCount":"24200"}, 
    ...
    "comments": [], 
    ...
}
\end{lstlisting}\vspace{-0.15in}
\caption{\centering \textbf{Video Metadata.} \textit{Metadata associated with each TikTok video contains popularity metrics, creation time, etc. }}\vspace{-0.15in}
  \label{fig:videometadata}
\end{figure}

\begin{figure}[htbp]
    \centering
    \begin{subfigure}[t]{0.49\linewidth} 
        \centering
        \includegraphics[width=\linewidth]{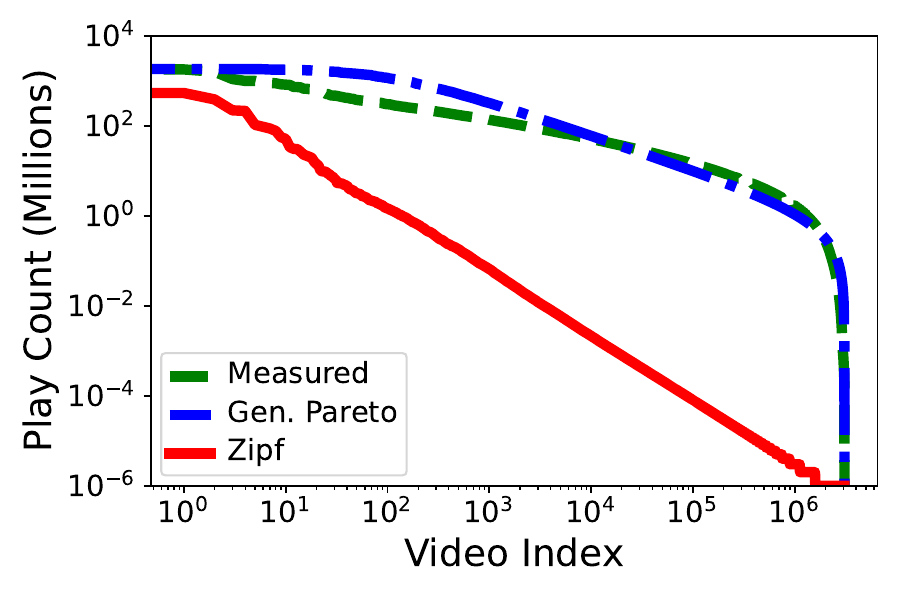} 
        \caption{\centering \textit{Video popularity follows the Generalized Pareto distribution.}}
        \label{fig:genpareto}
    \end{subfigure}
    \begin{subfigure}[t]{0.49\linewidth} 
        \centering
        \includegraphics[width=\linewidth]{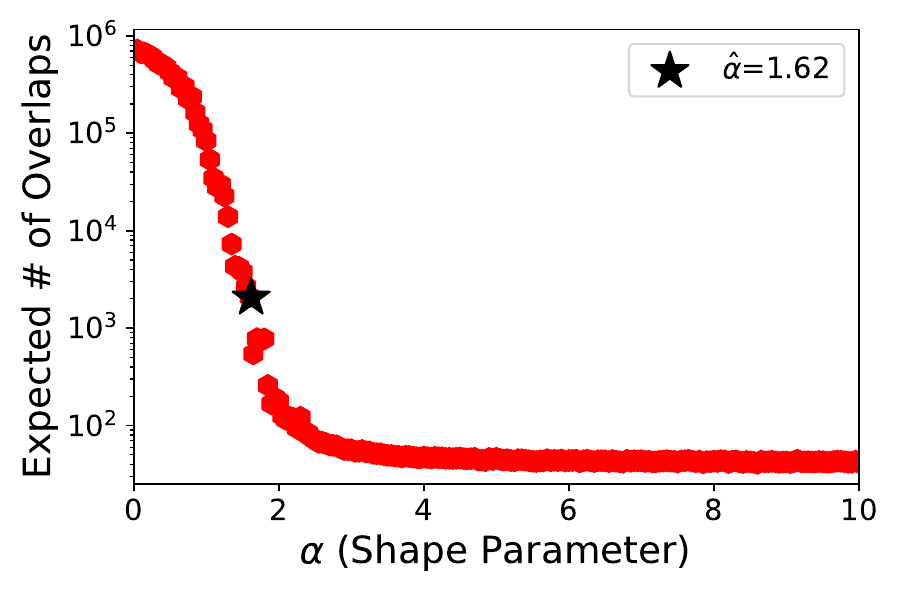}\vspace{-0.1in}
        \caption{\centering \textit{Expected number of overlaps between two users sampling from Pareto distributions.}}
        \label{fig:simulationanalysisonly}
    \end{subfigure} \vspace{-0.1in}
    \caption{\centering \textbf{Popularity Distribution in Short Videos.}}\vspace{-0.15in}
    \label{fig:mainfig}
\end{figure}

TikTok videos follow a highly skewed popularity distribution. We use our data donation exercise (\S\ref{sec:userstudy}) to collect and analyze the associated metadata of 2.65M unique videos. An example metadata response is shown in Fig.~\ref{fig:videometadata}.

We plot the play counts of the videos in our dataset in Fig.~\ref{fig:genpareto}, sorted by video popularity (most popular videos to the left). As shown in the figure, this distribution is extremely skewed, i.e., a small fraction (20\%) of the videos account for most (80\%) of the views. This distribution closely mimics the Pareto distribution \cite{castillo1997fitting}, which is popularly used for quantifying income inequalities \cite{pareto1896}. 

The short video popularity distribution is more skewed than the Zipf distribution (which is a straight line on a log-log plot) that is typically observed in popularity of webpages \cite{breslau1999web} and data center traffic \cite{zipfcontainer,jiang2020characterizing}. However, it matches the popularity trends of user-generated content in YouTube \cite{cha2007tube}. We are the first to observe this distribution in short videos.

{\para{Implication: }The highly skewed workload implies that, compared to traditional workloads, there is a larger opportunity to improve midgress utilization through intelligent caching.}

\vspace{3pt}\noindent{\it Observation 2: There is a larger opportunity to improve midgress utilization because of the highly skewed workload.}\vspace{3pt}

%% file: design.tex
\subsection{SILC's Lookahead Aware Cache Eviction}\label{sec:llf}

To reduce midgress traffic utilization and miss rate for a CDN serving short videos, \name\ exploits cross-user and temporal overlap in short videos served to different users. 

As discussed before, video popularity in TikTok follows a skewed Pareto distribution. The Pareto distribution is defined as follows:
\begin{equation}
f(x) = 
\begin{cases} 
\frac{\alpha x_m^\alpha}{x^{\alpha+1}}, & x \geq x_m, \\
0, & x < x_m,
\end{cases}
\label{eq:pareto_pdf}
\end{equation}
where $\alpha > 0$ is the shape parameter and $x_m$ is the scale parameter (minimum value of $x$ which is 1 in our case). A smaller $\alpha$ denotes a more skewed distribution. The observed Pareto distribution fits Eq.~\ref{eq:pareto_pdf} with $\alpha=1.62$.

We quantify the relationship between $\alpha$ and cross-user overlap using a simulation  with 100 users each sampling 150 videos from this distribution. The results are averaged across 100 runs and plotted in Fig.~\ref{fig:simulationanalysisonly}. The plot confirms the intuition that overlap across users increases with decreasing values of $\alpha$. Note that the $\alpha$ in short video popularity is small enough (or the skew is large enough) that it lands on the upward sloping part of this curve. The high level of skew denotes an opportunity to extract significant caching benefits.

\begin{figure}
    \centering
    \includegraphics[width=.3\textwidth]{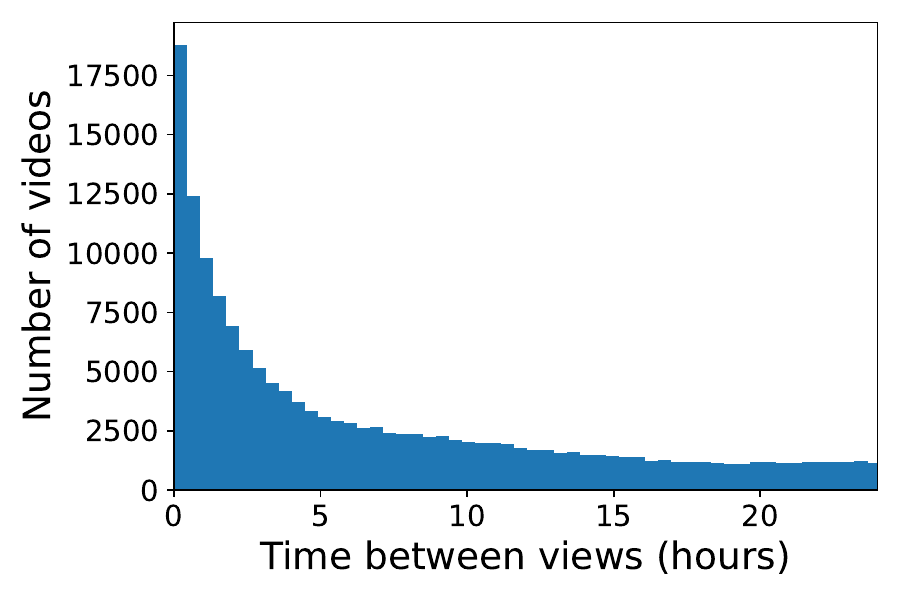}\vspace{-0.15in}
    \caption{\centering \textbf{Distribution of Time between Successive Views.} \textit{23\% of videos in our dataset were watched by more than 1 person, of which 54\% were watched within 24 hours.}} \vspace{-0.15in}
    \label{fig:bh_overlap}
\end{figure}

We also observe significant {\it spatial} (cross-user) overlaps in our empirical data. We quantify the overlap in videos watched by 100 users over a span of 6 months.  We plot the number of videos watched by more than one user and the distribution of time-gap between their viewing times in Fig.~\ref{fig:bh_overlap}. Across our dataset, we observe that 23\% of the videos were watched by more than one user, of which 54\% were watched within 24 hours of each other. The plot also suggests high temporal locality in user interests (i.e., the number of videos with distant overlapping views are fewer). Note that, this overlap is observed in just 100 users. A CDN typically serves tens of thousands of users (since there are 1B+ TikTok users and Akamai has 4200 sites~\cite{DanAkamai}). Therefore, we expect the temporal and geographical overlap in video content to be higher in practice. 

\vspace{3pt}
\noindent\textit{Observation 3: Skewed distributions lead to larger cross-user overlaps.} 
\vspace{3pt}

\para{Least Lookahead Frequency for Cache Eviction: }To exploit the larger cross-user overlaps, we design a new cache eviction policy. Cache eviction is a fundamental problem in CDN design. A CDN can store a limited number of videos and must regularly evict videos when new videos need to be added to the cache. Optimal eviction policies typically require prior knowledge of the entire sequence of future requests or an infinite cache, making them impractical.

Yet, short video streaming  offers a unique opportunity to leverage limited (myopic) knowledge of future videos. 
Because each user receives a manifest file (generated by the recommendation algorithm) from the CDN, the CDN can \textit{lookahead} into the future. However, this lookahead is {\it myopic}, because the manifest file contains a finite set (say 30) of videos. 

Based on this insight, we design the \textit{Least Lookahead Frequency (LLF)} cache eviction policy. Intuitively, LLF evicts videos with the least frequency in the limited lookahead available to the CDN. This is unlike traditional policies which use past as a proxy for the future and use historical decisions for making new ones. Using LLF in \name, we can look into the future and use the future access patterns to make decisions.

\begin{figure}[htbp]
    \centering
    \begin{subfigure}[t]{0.49\linewidth} 
        \centering
        \includegraphics[width=\linewidth]{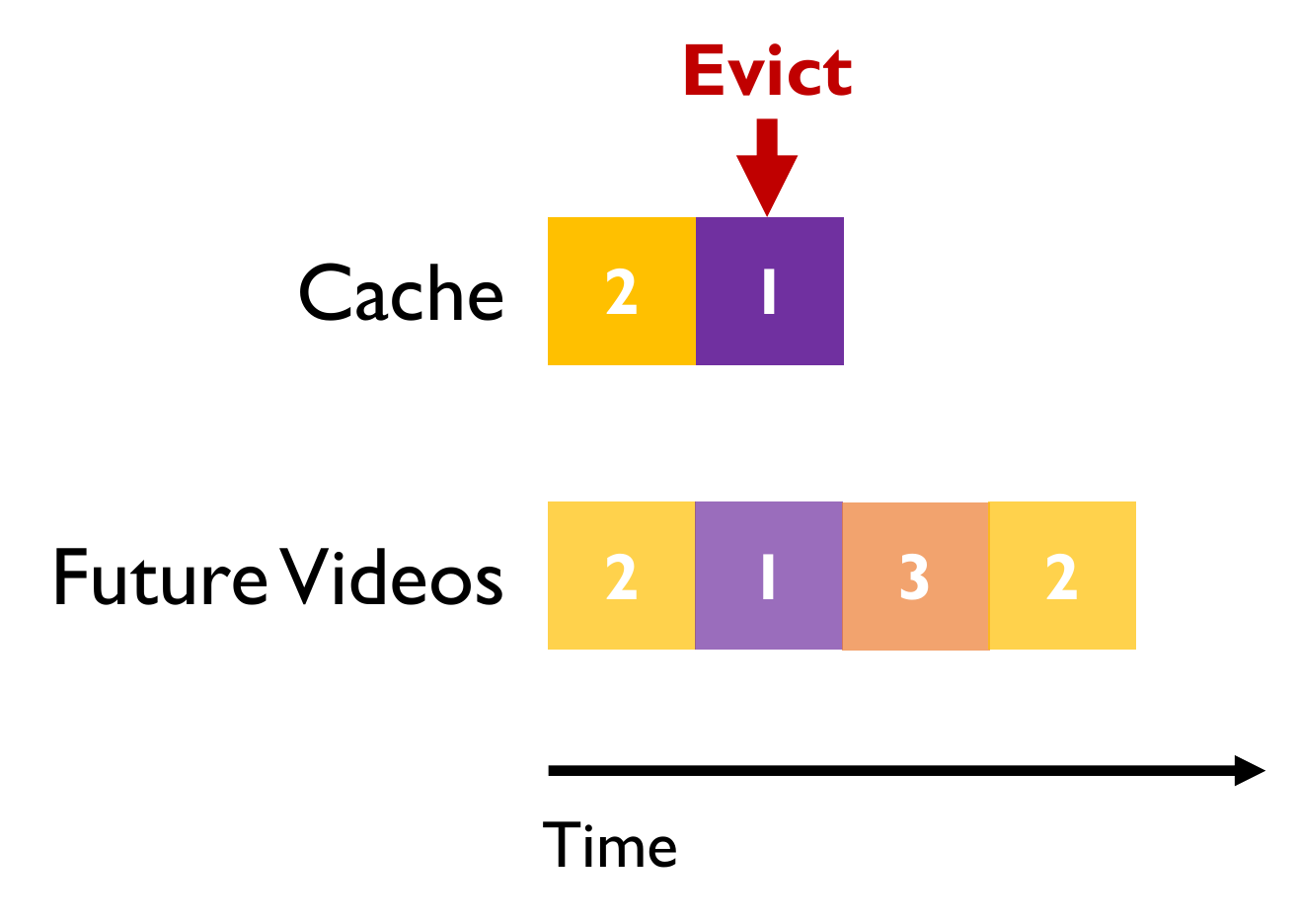} 
        \caption{\centering \textit{\name's LLF cache-eviction policy evicts videos with the least views in the future requests.}}
        \label{fig:eviction}
    \end{subfigure}
    \begin{subfigure}[t]{0.49\linewidth} 
        \centering
        \includegraphics[width=\linewidth]{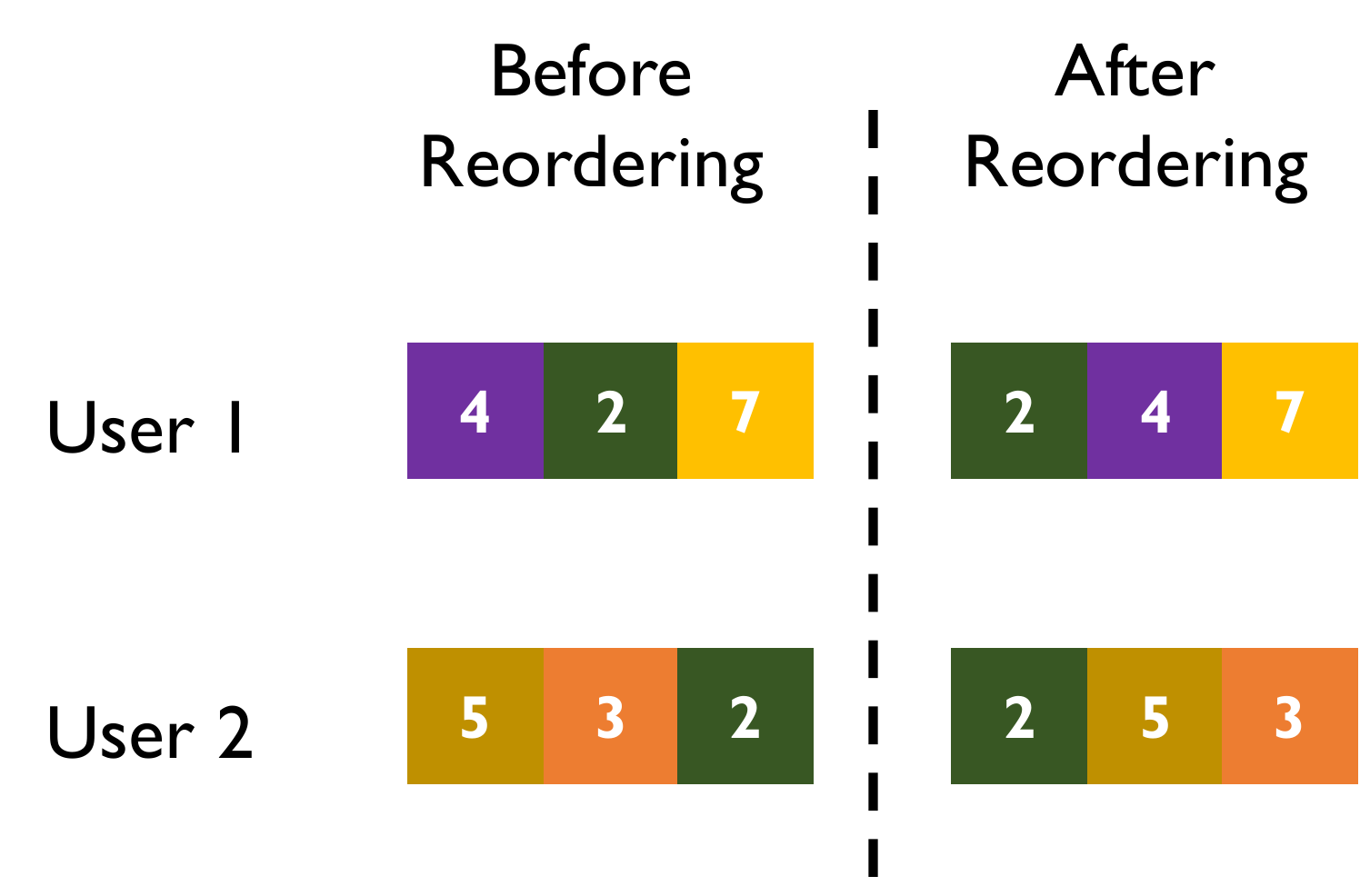}
        \caption{\centering \textit{\name\ reorders manifest files to maximize temporal overlap across users to reduce midgress traffic.}}
        \label{fig:reordering}
    \end{subfigure} 
    \vspace{-0.1in}
    \caption{\centering \textbf{Example of \name's Components.}} \vspace{-0.1in}
    \label{fig:mainfig}
\end{figure}

To define our proposed eviction policy formally, assume that the cache has $K$ videos $\{v_1,v_2,...,v_K\}$. The CDN computes the future frequency, $f_i$, for each video by counting $v_i$'s occurrences that have not yet been served to the client (across manifest files of all users that the CDN is serving at the moment). On every new manifest file seen, the CDN increments the associated future frequencies $f_i$. Future frequencies decrease  as videos are served. If the CDN needs to evict a video from cache, it  evicts $v_k$ such that \begin{align}
k=\arg\min_{i\in\{1,..,K\}} f_i
\end{align}
If there is a tie, LRU is used as tie-breaker. Fig.~\ref{fig:eviction} shows an example of an eviction driven by LLF.

\subsection{SILC's Manifest File Reordering}
\label{sec:reordering}

We make two further observations to further improve \name's caching policy. 

\vspace{3pt}\noindent{\it Observation 4:} {\it Cache efficiency is improved if multiple users fetch the same video around the same time.}\vspace{3pt}

Essentially, if all users fetch the same video around the same time, the CDN will need to fetch this video once from the content provider. Once this video is served to all users, if needed, it can be removed from the cache, making room for new videos to be cached. 
In general, increased  spatio-temporal overlap improves cache hit rate and reduces midgress by reducing the time a video spends in the cache. 

\vspace{3pt}\noindent{\it Observation 5 (Hypothesis): Users are engaged by the \textit{content} rather than the \textit{order} of the videos.}\vspace{3pt}

We empirically validate this hypothesis in \S\ref{sec:userstudy}. Observation 5 opens up a key design space in \name---that the {\it set} of videos {\it within} a manifest file all need to be served, but the {\it order} in which they are served can be modified by \name. Note that this reordering is limited to the small set videos within a given manifest file, and does not extend across manifest files (even for a given user). \name\ also neither drops videos from nor includes videos into a given manifest file, thus adhering to  recommendation-agnosticity. 

Based on these observations 4 and 5, we propose a new technique to improve cache design: {\it online manifest file reordering}. Online manifest file reordering occurs at the CDN. It takes each currently active manifest file (i.e., users currently receiving videos), and reorders the videos in this list based on already available future information. Manifest file reordering attempts to maximize the time overlap among the resultant manifest files. 

{Fig.~\ref{fig:reordering} illustrates benefits of reordering with a simple example involving two users. Assume all videos are equally long,  user latencies are similar,  users view all videos, and the CDN cache size = 1 video. The left (Before Reordering) has a midgress of 6 videos (all fetched by CDN), while the right side (After Reordering) has a midgress of 5 videos. 

Note that, the manifest files only define the order of the videos and not the exact times they will be watched by a user. A user may skip, replay or partially view videos. Therefore, it is not possible to synchronize the same video exactly across multiple users. However, to gain benefits from reordering, we only need multiple accesses to a single video to be close in time. As they get closer in time, the time spent by each video in cache decreases and the caching efficiency increases.

To utilize these insights across multiple users, \name\ uses two techniques to reorder videos within  currently-active manifest files at the CDN. The first is reordering the videos in each active manifest file such that the videos with the maximum cross-user overlap (i.e., the video is in the active manifest files for the largest subset of users) are placed at the beginning of that manifest file.

The second reordering technique places videos which are already in the CDN cache at the beginning of the manifest files. To maximize spatio-temporal, each reordered manifest file contains (1) videos in the manifest file that are also present in the cache (in descending order based on the frequency count across users), followed by (2) videos that are not in the cache but are expected to be requested and finally, (3) the remaining videos in the order they appeared in the original manifest file. {This reordering is done in real time and the processing overhead is low as the CDN keeps track of frequencies as it receives manifest files and serves videos.}

\subsection{Time Complexity of SILC} \label{sec:timecomplextiy}

We analyze  time complexity of different \name\  components.

\para{Time Complexity of LLF:} \name's cache is implemented using 2 hashmaps and a min-heap. The first hashmap is used to keep track of lookahead frequencies of videos in the cache. The second hashmap does the same for videos outside the cache. A tuple containing lookahead frequency, time of access and video is inserted into the min-heap, such that the lookahead frequency is used to sort the tuples. 

\squishlist

\item \textbf{Search: } To check if a video is in cache, the cache hashmap can be used to perform an O(1) search.

\item \textbf{Insert: } To insert a video in cache, we need to update the in-cache hashmap and the heap. In a hashmap, inserts are O(1). In the heap, inserting a video such that videos with minimum future frequency (least requested videos in the lookahead future) are available to be accessed readily requires O(log\textit{n}) time.

\item \textbf{Evict: } Removing a video from the min-heap requires O(log\textit{n}) time. In the hashmap, eviction of the key of the video evicted from the min-heap can be done in O(1). 

\item \textbf{Updating Future Frequencies: } As manifest files are received at the CDN, both the in-cache and out-of-cache hashmaps are updated. The number of modifications is equivalent to the number of videos in a manifest file (30 in our experiments, but generally a much smaller number than the total number of videos).
\squishend

Overall, \name's cache serves requests with a time complextiy of O(log\textit{n}). A cache like LRU can serve requests in O(1) time. Despite the extra overhead of LLF incurred by keeping track of frequencies, the response latency of \name\ does not differ much from that of any low overhead caching policy like LRU. For a trial of the same workload, the median response latency of \name\ (2.54 $\mu$\text{s}) is 2.67 times worse than LRU (0.95 $\mu$\text{s}). The extra time in response is still much smaller than the time between consecutive requests.

\para{Time Complexity of Reordering Manifest Files}
Our network analysis of TikTok traffic reveals that manifest files are already sent to users through the CDN. Thus, online reordering of videos does not incur any extra network latency. \name\ reorders every new manifest file received at the CDN based on the frequencies of future requests and availability of objects in cache and sends it back to the client. While sorting a list has a time complexity of O(\textit{n}log\textit{n}), because the size of the manifest file is very small (30 videos in our experiments) and users request manifest files at different times, we expect this to not be an overbearing overhead.

%% file: user-study.tex
\vspace{-5pt}\section{User Study}\label{sec:userstudy}
We conduct a hybrid user study involving both in-person participants on a university campus in United States, and online (global) participants via Prolific. The former in-person study follows the principles of {\it within-subjects \cite{withinsub} A/B testing} (also known as Comparative Usability Testing)~\cite{comparativetesting}. The latter follows the principles of a data donation program~\cite{donation}.
Our user study has 44 participants and we received data donations from 100 TikTok users. Each study took between 30 minutes to 1 hour of participant time and 2-3 hours of researcher time (including advertising, scheduling, prep, actual study, post-mortem analysis to ensure we had enough data to analyze). 

We use the study to create a large dataset of short videos that forms the foundation of our population analysis (\S\ref{sec:analysis}) and facilitates our evaluation of \name\ using real workloads (\S\ref{sec:results}). We also use this user study to test the impact of reordering manifest files on user experience. Given the interactive nature of short videos, 
we evaluate whether reordering of videos will affect user's Quality of Experience (QoE). Our user study was reviewed and approved by our Institutional Review Board (IRB) for ethics and privacy. Overall, our study aims to answer the following research questions (RQs):

\vspace{-0.1cm}
\begin{tcolorbox}[colback=red!5!white,colframe=black!75!black]
\vspace{-0.2cm}
\squishlist
\item {\bf \underline{RQ1}}: Can the popularity of short videos, measured by the play count, be accurately modeled using a Pareto distribution?
\item {\bf \underline{RQ2}}: 
 Does reordering of videos affect user engagement in TikTok-style short video applications? 
\squishend\vspace{-0.1in}
\end{tcolorbox}
\vspace{-0.1cm}

\subsection{RQ1: Curating a Short Video Dataset}
\label{sec:datadonation}
The first part of our study consisted of a Data Donation program, similar to that conducted in \cite{donation}. This program ran between November 2023 and September 2024. Users that had an active TikTok account and regularly engaged with the platform could choose to participate in the study and share their TikTok video browsing history. The video browsing history contains a list of videos seen by the user (plus the time they were seen, the watch length, and other metadata information) over the span of approximately 6 months and could be requested to download via TikTok. In exchange for their anonymized browsing history, users were compensated \$15. We enlisted participants for our study by using posters and social media announcements.

{100 participants opted in to our Data Donation program, and shared their TikTok browsing history with us.  67\% were female and 32\% were male. 60\% of our participants were between the ages of 18 and 24, 36\% were between 25-34 and 4\% were between 35-44. In terms of ethnicity, 44\% identified as White/Caucasian, 26\% as Asian, and 14\% as Latino, with the remaining participants representing other ethnic groups. Additionally, 72\% of participants were from North America, 20\% from Europe, 5\% from Africa and the remaining were from Asia and South America. Overall, each participants's shared browsing history spanned a duration of 170 to 450 days, representing the most recent activity of each user. The median number of unique videos per browsing history was 46,600. On average, each user saw 332 videos in a day.

The resultant 100 browsing histories contained 3.9M unique videos, of which we could collect metadata for 2.65M videos (some videos had been deleted, removed, or were inaccessible when we checked them for metadata). Some key aspects are as follows: 

\squishlist
\item Fig.~\ref{fig:bh_popularity} shows that the popularity statistics (play count, like count, share count, and comment count) follow the Pareto curve (already discussed in \S\ref{sec:analysis}).

\item The median video duration is 23s, the 25th and 75th percentile are 11s and 60s respectively. 75\% of videos are under a minute, and 92\% videos are under 2 minutes.

\item 43\% of the videos are less than 3 MB, and 12\% of the videos are under 1 MB. 78\% of the videos are under 10 MB. The smallest video is 11 KB. The largest video is 1000 MB.
\squishend

The play counts of videos collected from this dataset confirm that the popularity of videos in short video streaming apps follows the Pareto distribution. This further validates the high probability of overlap between users' viewed videos.

\begin{figure*}
\begin{subfigure}{0.23\textwidth}
\centering
\includegraphics[width=1\linewidth]{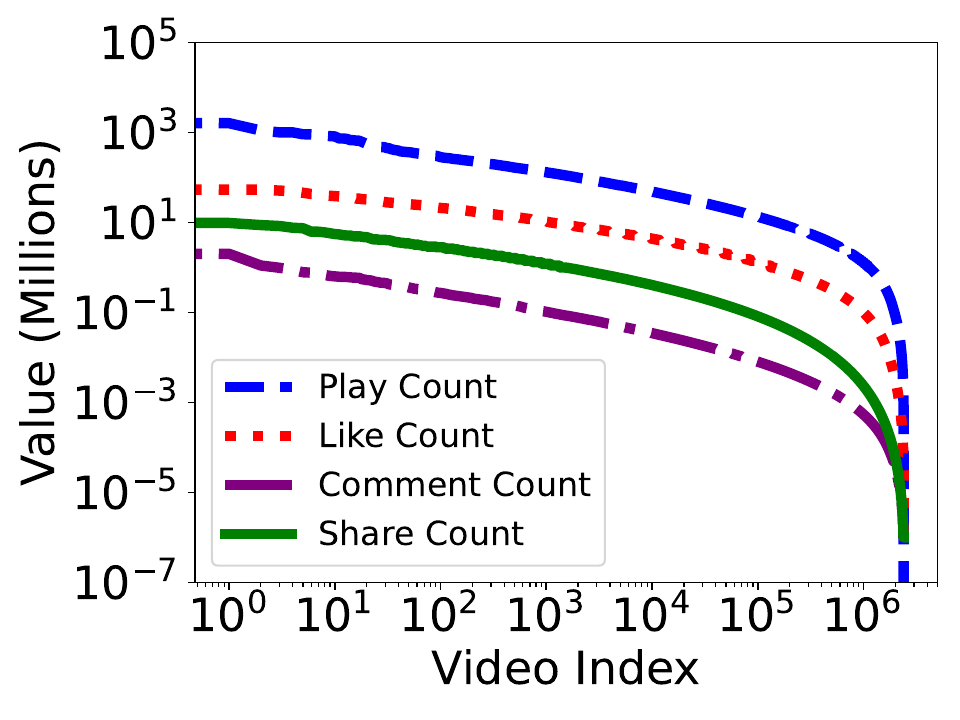} \caption{\centering \textit{Popularity of Videos collected via the Study}}
\label{fig:bh_popularity}
\end{subfigure}\quad
\begin{subfigure}{0.23\textwidth}
\centering
\includegraphics[width=1\linewidth]{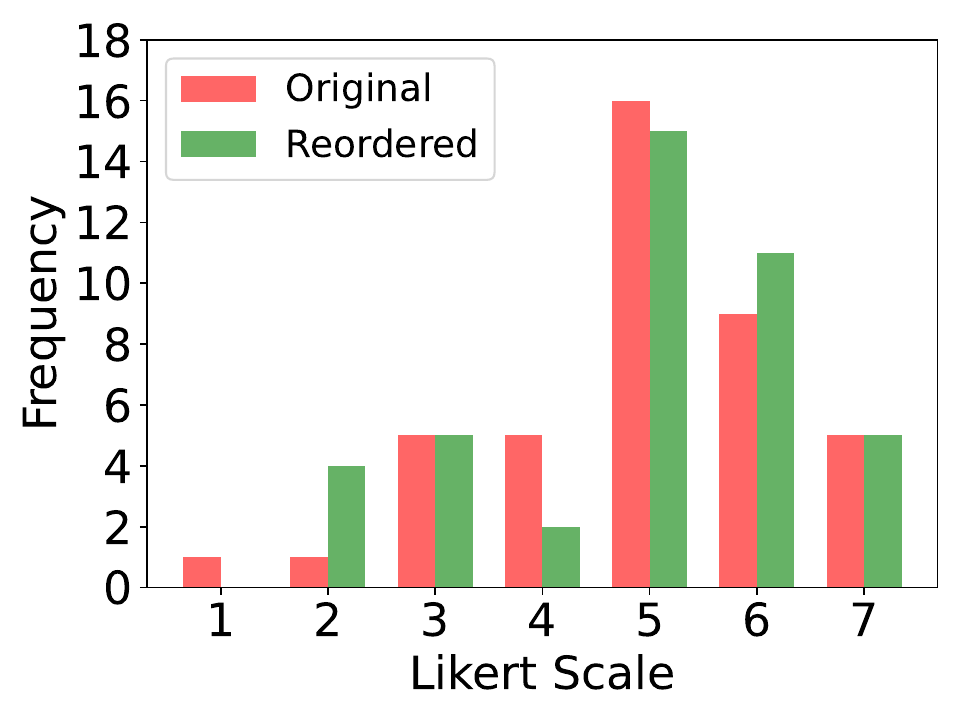} \caption{\centering \textit{User Engagement Scores for all Sessions}}
\label{fig:reorder_impact}
\end{subfigure}\quad
\begin{subfigure}{0.23\textwidth}
\centering
  \includegraphics[width=1\linewidth]{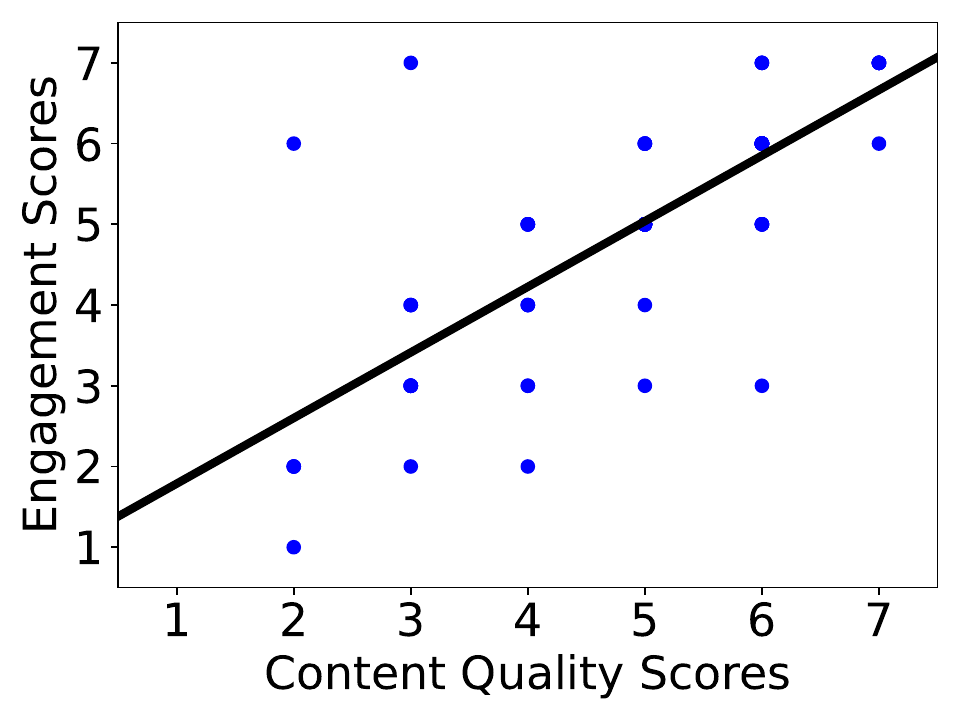}
  \caption{\centering \textit{Content Quality vs Engagement}}
  \label{fig:content_eng}
\end{subfigure}\quad
\begin{subfigure}{0.23\textwidth}
\centering
  
  \includegraphics[width=1\linewidth]{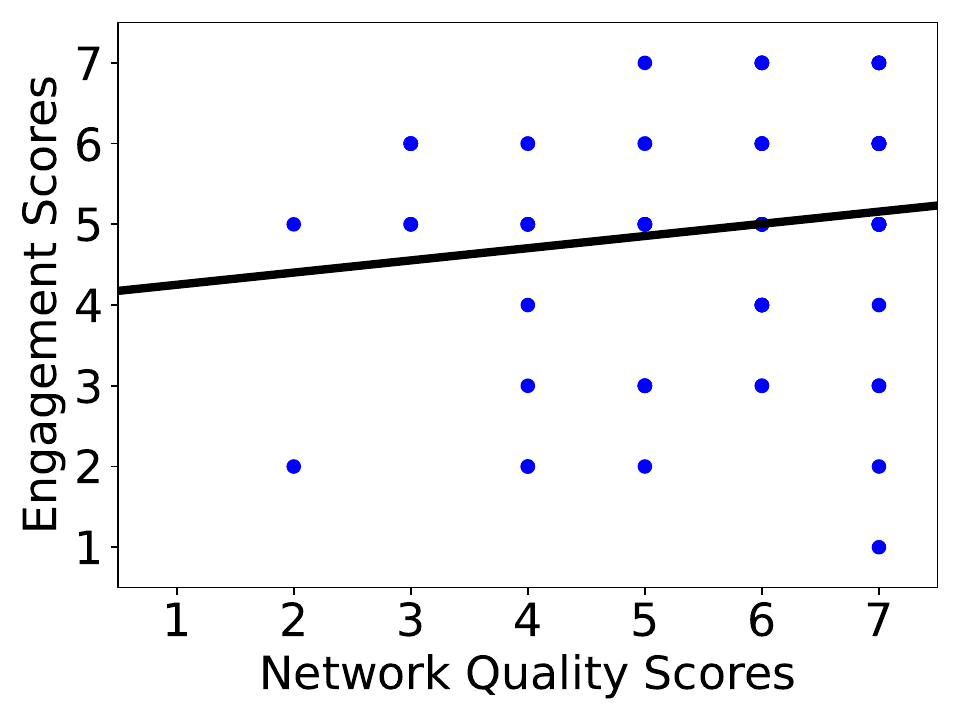}
  \caption{\centering \textit{Network Quality vs Engagement}}
  \label{fig:network_eng}
\end{subfigure}\quad
\vspace{-0.1in}
\caption{\centering \textbf{Metrics Collected in the User Study.}}\vspace{-15pt}
\label{fig:bh_metrics}
\end{figure*}

\subsection{RQ2: Impact of Reordering on User QoE (and Other Insights)}

In the in-person study, we had 44  participants (subset of the 100 participants that shared their browsing histories). The study duration was 45 minutes, and participants were remunerated with \$15. The in-person study was conducted between November and December 2023.

The study uses within-subjects A/B testing (also known as Comparative Usability Testing) and comprised of two 10-minute sessions for each participant. In both their sessions, a user logged into their own TikTok account using a web browser, and viewed videos on their TikTok ``For You Page (FYP)'', just as they normally would. 
In both the sessions, the content shown to participants  came directly from their own TikTok account, i.e., from the TikTok recommendation algorithm---the content of the manifest files and the resultant videos was not under our study's control. 

For each user, one of the two sessions was unmodified, i.e., the control session. For the other session, we showed a random permutation of the videos linked from each manifest file. 52\% of users received the unmodified session first followed by the modified (control) session, and the remaining users received them in the opposite order. We did not tell users which session was modified and which was control. To enable this modifications we implemented a browser extension atop Chrome browser. This extension  received manifest files and modified them invisibly, in real time, and in the background. We used the browser extension for both settings---control (unmodified) and reordered. 

After each session, the participant was asked to rate Session 1 and Session 2 separately on a Likert scale (1-7)---the participant rated  user engagement, content quality, and network stability. 
 
All questions are listed in the appendix for completeness.

\para{\underline{Key Finding}}: We find that (1) Video reordering does {\it not} affect user engagement metrics, (2)  Users  observe lower engagement when the perceived network quality drops (e.g., glitchy sessions).  

Fig.~\ref{fig:reorder_impact} shows user engagement scores for both {\it Original} (control, unmodified) session and {\it Reordered} session. We use four diverse  statistical tests to analyze whether the two distributions match each other or are different from each other. (1) The mean (and standard deviation) engagement score for the \textit{Original} sessions is 4.92 (1.36) and \textit{Reordered} sessions is 4.92 (1.45). (2) The median engagement score received for both the sessions was 5. (3) The Wasserstein distance (Earthmover's distance) \cite{Wass24}  between the Original and Reordered distributions was 0.19, which is small enough to indicate the distributions are similar. {(4) Finally, we used the Wilcoxon signed-rank test \cite{wilcoxon}, a non-parametric test for paired ordinal data, to assess differences in user ratings between the two sessions. The results showed no significant difference between the two sessions (W = 367.5, p = 0.965), indicating that difference in engagement scores was not statistically significant.}  Together all four tests validate the hypothesis that reordering videos (inside a manifest file)  does not affect user engagement.

\para{Other Insights from User Study: }
\squishlist
    \item \textit{Content drives engagement: }As the content quality scores by a user goes up, a user's engagement score also goes up (Fig.~\ref{fig:content_eng}). This reinforces our hypothesis that video engagement is driven by \textit{content}, rather than \textit{order}.
    \item {\textit{Network conditions are important: }If a user perceives a laggy network or experiences glitches, the engagement score falls  (Fig.~\ref{fig:network_eng}). This indicates that {\it CDN caches are important for lower delays and improved network performance}.}
    \item \textit{Overlap across users: } In spite of TikTok having billions of users, even at our scale of 44 users watching TikTok for just 20 minutes (across two weeks), we found that 93 videos were  watched by more than one user (and some by even four). This confirms our hypothesis that {\it there is overlap in the content consumed by users of a push-based recommendation system like TikTok}.
\squishend

%% file: eval-2.tex
\vspace{-10pt}
\section{Evaluation of \name\ Performance}
\label{sec:results}

Our evaluation of \name\ quantifies the benefits of our design choices and addresses the following questions:

\squishlist
\item How much does \name\ reduce midgress traffic in comparison to existing heuristic and learning-based cache eviction strategies (\S \ref{sec:midgress} and \S\ref{sec:hitrate})?
\item How do \name's benefits change with cache size (\S \ref{sec:ablation})?
\item How does \name\ compare to caching all popular videos (\S \ref{sec:ablation})?
\item How much does manifest file reordering in \name\ contribute to its cache miss improvement (\S \ref{sec:ablation})?

\squishend

\subsection{Evaluation Setup and Baselines}

Our evaluation consists of two components: an emulation at scale (\S\ref{sec:emulation}) and a simulation to compare with 10 other caching baselines (\S\ref{sec:simulation}).

\subsubsection{Emulation}
\label{sec:emulation}
To mimic a real-world setting, we emulate 
10,000 users each viewing 150 videos.  
Our  setup contains: (a) a TikTok origin server, (b) a CDN cluster consisting of 10 homogeneous servers, and (c) a client server that randomly spawns multiple new users who download manifest files and videos through the CDN. Each experiment runs on 3 separate machines from the Cloudlab cluster \cite{duplyakin2019design} with a link capacity of 10 Gbps. The default CDN cache size for each server is realistically set at 10GB each for a total of 100GB at the CDN cluster. We also evaluate \name\ at different cache sizes (\S \ref{sec:ablation}). These sizes were calculated by scaling down  Akamai's CDN cluster sizes to one traffic class (we consider short videos as one traffic class) \cite{sundarrajan2020midgress}.

Our emulation also mimics how clients today connect to Tiktok. Each client requests manifest files from our server. Clients then request every videos one by one from a server in CDN cluster (based on the object's sha256 hash). Once 10 videos in a manifest file remain, clients restart the process by requesting for a new manifest file for a total of 5 times. We start by running a random subset of 10 clients. Then, every 0.01 s, we add more clients---we  max out at 500 clients. When a client leaves, a new one joins.  
At the end of each emulation run, we obtain a serialized interleaving of video requests from multiple emulated users served at each server within the CDN cluster. 

\subsubsection{Simulator with Baselines}
\label{sec:simulation}

We compare \name\ against multiple cache eviction strategies. Concretely, we extract traces from the emulation above, and use a publicly available caching simulator \cite{lrb}. We compare against two categories of cache eviction: Heuristic (total 6 strategies), and Learning-based (total 4 strategies). We also compare against our implementation of Belady's MIN algorithm~\cite{beladymin}.

\para{(a) Heuristic}: We compare against these heuristic eviction policies: \textbf{LRU} \cite{lru}, \textbf{LFU} \cite{lfu}, \textbf{FIFO} \cite{fifo}, \textbf{GDSF} \cite{gdsf} and \textbf{LFUDA} \cite{lfuda}. We also compare with \textbf{Random} eviction. Except Random, each of these uses a combination of recency (LRU, FIFO), frequency (LFU, LFUDA) and size (GDSF) to decide the object that will be evicted from the cache. More detailed information is in the appendix 
(Table \ref{table:heuristic}).

\para{(b) Learning-Based} We compare against: 
\textbf{LeCaR} \cite{lecar} (intelligently selects between LRU and LFU), \textbf{LRB} \cite{lrb} (approximate's Belady's MIN algorithm \cite{beladymin} and predicts  farthest request in a selected window), \textbf{LHD} \cite{lhd} (predicts hit density of objects) and \textbf{AdaptSize} \cite{adaptsize} (uses size to admit objects). We fine-tune LRB and AdaptSize to our workloads and use default configurations for LeCaR and LHD (changing parameters did not impact  results). More details are in the appendix. 

We also implemented a practical version of Belady's MIN algorithm that we call Furthest-in-Future (\textbf{FIF}). FIF evicts the video which is the farthest expected to be used in the future based on the limited lookahead at the CDN provided by the manifest files. 

\subsection{Emulating Users with Distinct Personalities and Drift}\label{sec:manifestgen}

To evaluate \name\ at scale, we need to emulate thousands of users. We emulate these users using different combinations of the 100 participants from our user study (\S\ref{sec:datadonation}). Each emulated user needs to have both (i) a unique personality, and (ii)  temporal drift in interests.

\para{\textbf{User Personality: }}
 
We model each user around a combination of $K$ randomly sampled participants from our 100 real participants. $K$ is set uniformly at random in $[5, 100]$. Each emulated user watches videos chosen from the browsing history of those $K$ participants that are assigned to it.

\para{\textbf{Temporal Drift: }}
Since we have long-term traces for each of our real participants, we need to decide which portion of each user trace to extract for each generated user.  We extract this is in a finite time window with start time $t_s$ and an end time $t_e$. 

This window is set to 4 days for our experiments. We generate users in batches of 100, with each batch shifting the starting window in increments of 1 day. This models temporal drift.

\para{Sampling Videos: }
We select 150 videos to be watched by each emulated user. The probability of selecting video, $v$, is: 
\vspace*{-0.2cm}
\begin{align}
    P(v) = \beta \frac{C_v}{\sum C_v} + (1-\beta)\frac{1}{N_v}
\end{align}
where $C_v$ is the number of views for video $v$ and $N_v$ is the total number of videos.
This probability is a combination of two factors. The first term increases the likelihood of more popular videos to be picked in proportion to their play counts (modeling the Pareto distribution). 

The second term uniformly samples across all videos. 

We use the parameter $\beta$ to vary the contribution of the two terms. $\beta=1$ samples based on popularity alone, with lower values of $\beta$ making the distribution more uniform. We vary $\beta$ to model different overlap degrees across users - as $\beta$ increases from 0 to 1, the overlap between videos seen by users increase. 

\begin{figure}
  \centering
  \includegraphics[width=0.8\linewidth]{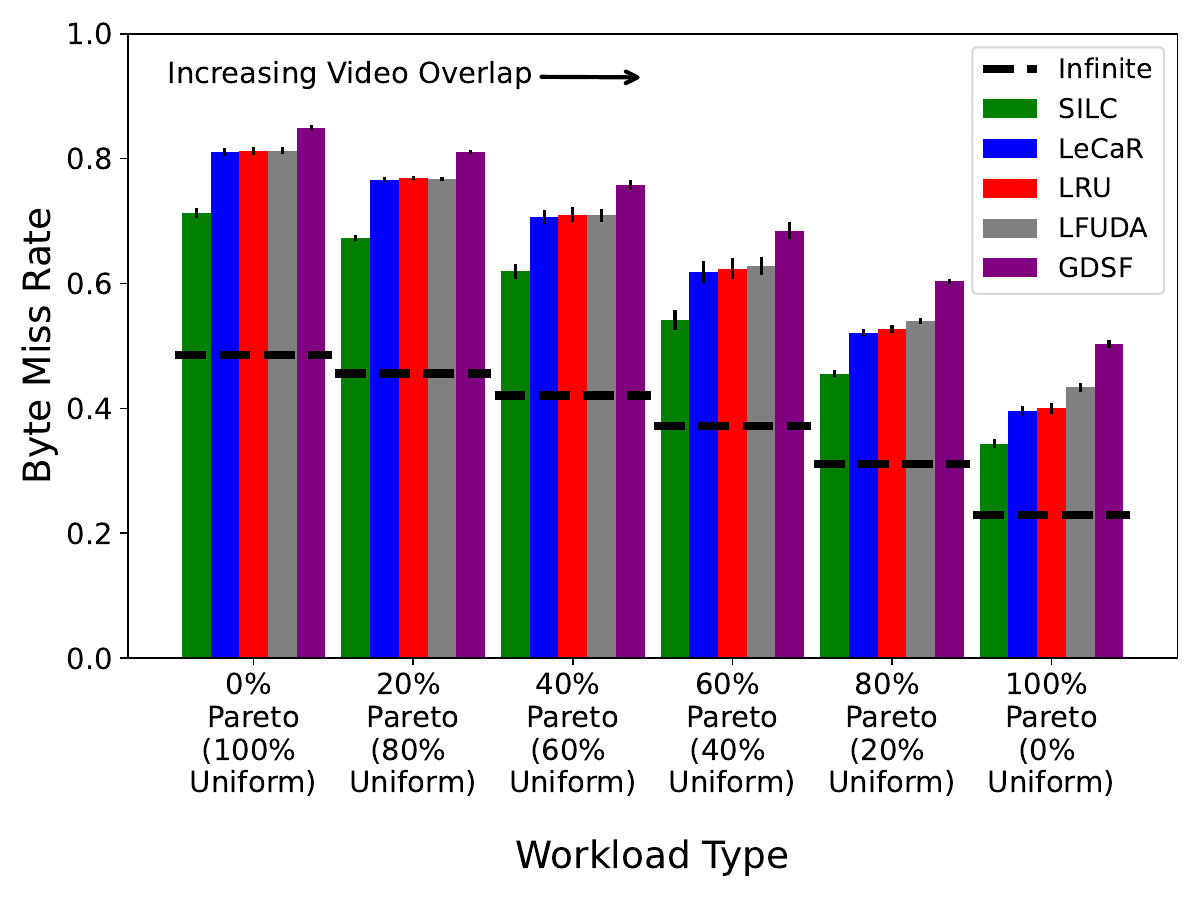} 
  \vspace{-0.1in}
  \caption{\centering \textbf{Byte Miss Rate.} \textit{Byte miss rate of \name\ compared to the best learning based (LRB), recency (LRU), frequency (LFUDA) and frequency + size (GDSF) heuristic eviction policies. \name\ outperforms all baselines by at least 11.1\%. The dotted line refers to the best rate possible with an infinite cache size. }}
  \label{fig:midgressPercent}\vspace{-0.2in}
\end{figure}

\begin{table*}
\includegraphics[width=\linewidth]{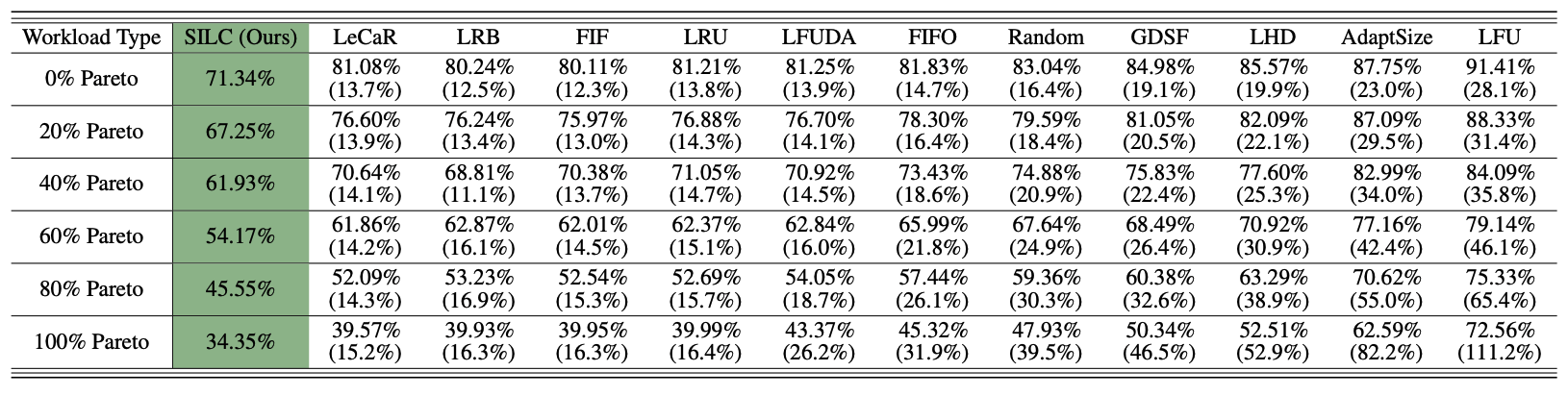}
\vspace{-0.1in}
\caption{\centering \textbf{Byte Miss Rate.} \textit{Byte miss rate comparison for all workloads and baselines. Values in brackets indicate percentage difference with respect to \name. 
}}
\label{table:bytemiss}
 \vspace{-0.2in}
\end{table*}

\begin{table*}
\includegraphics[width=\linewidth]{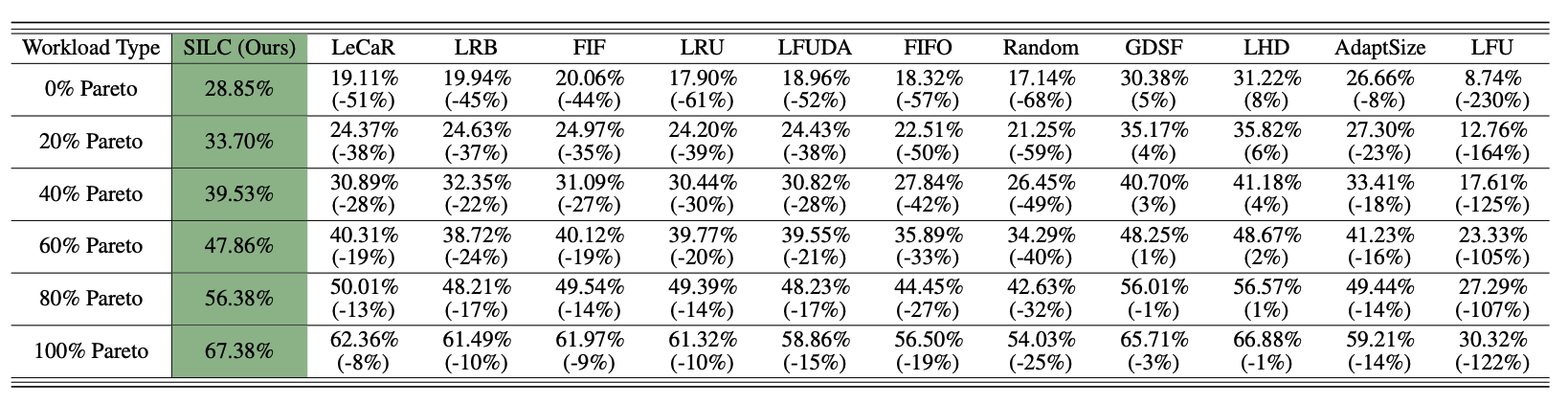}
\caption{\centering \textbf{Object Hit Rate.} \textit{Object hit rate comparison for all workloads and baselines. Values in brackets indicate percentage difference with respect to \name. 
}}
\vspace{-0.2in}
\label{table:objhit}
\end{table*}

\subsection{Midgress} \label{sec:midgress}

Recall that CDN midgress is the traffic between CDN and content provider, and is correlated with byte miss rate. 
Fig. \ref{fig:midgressPercent} plots the relationship between the different workloads and the byte miss rate. 
 
We only show the best-performing among the baseline heuristics. Table \ref{table:bytemiss} shows all baselines. 
Object miss rate is in appendix (Table \ref{table:objmiss}).  

\name\ outperforms all baselines on midgress traffic reduction. Across all workloads, other baselines increase midgress by at least 11.1\%. Some baselines, e.g. LFU at 100\% Pareto sampling, has a byte miss rate that is 111.2\% bigger than \name's (implying over 2$\times$ midgress compared to \name). This shows that \name\ can significantly reduce midgress costs at CDNs. 

As the percentage of Pareto sampling increases, so does video interest overlap across users, and thus all eviction policies perform better. 
A higher overlap
creates more opportunities for extracting higher caching benefits. Learning-based policies like LeCaR and LRB, despite performing better than heuristics, do not achieve performance close to \name, since they are unable to make good estimates about highly skewed distributions. \name\ on the other hand, makes informed decisions about eviction 
with correct knowledge of the near-future. 
We note that recency-based policies (LRU, LFUDA) do better than those that do not. 
 
This is because size and past frequency are not good proxies for the future in short video workloads---size does not say much about video popularity, and popular videos phase out over time. Also, \name\ has a 12.3\% improvement over FIF (our practical implementation of Belady's MIN algorithm). This is because even though future requests are present in the manifest files, the exact time and order of requests needed for Belady's algorithm can vary and needs to be approximated for it to do well.

\subsection{Cache Hit Rate} \label{sec:hitrate}
Cache hits represent the number of times a video was served from the cache. A cache hit means that the response to a user is quick and the CDN provider saves cost on midgress bandwidth. We measure cache hit in terms of objects in Table \ref{table:objhit} (byte hit rate is detailed in the appendix in Table~\ref{table:bytehit}).

\name\ performs better than almost all eviction policies, except  GDSF and LHD at lower Pareto percentage. Both GDSF and LHD optimize for ratios of popularity (measured in frequency) to object size. In short video systems, 
really popular videos have a high frequency to size ratio. Even though such popular short videos remain in the cache and can be packed together to achieve a higher object hit rate, they do not help with reduction of midgress traffic.  
We reiterate that in terms of the {\it byte} miss rate (directly correlated to midgress costs), \name\ outperforms GDSF and LHD by at least 
19.1\% to 52.9\%. 

\subsection{Microbenchmarks and Ablation}\label{sec:ablation}
In this section, we analyze various aspects of \name.

\begin{figure*}
    \begin{subfigure}{0.24\textwidth}
      \includegraphics[width=\linewidth]{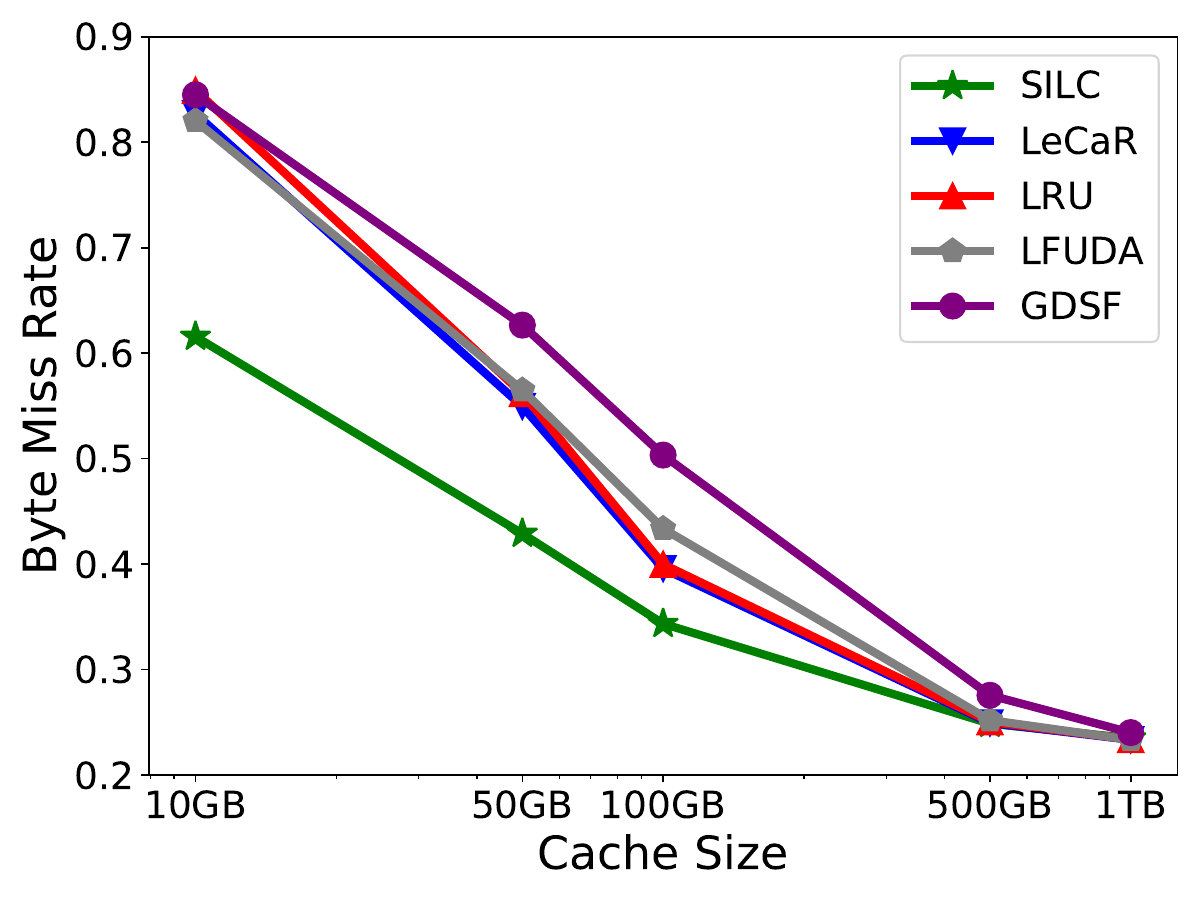} 
      \caption{\centering \textit{Impact of cache size}}
      \label{fig:cache40}
    \end{subfigure}\hfill
    \begin{subfigure}{0.24\linewidth}        
        \includegraphics[width=\linewidth]{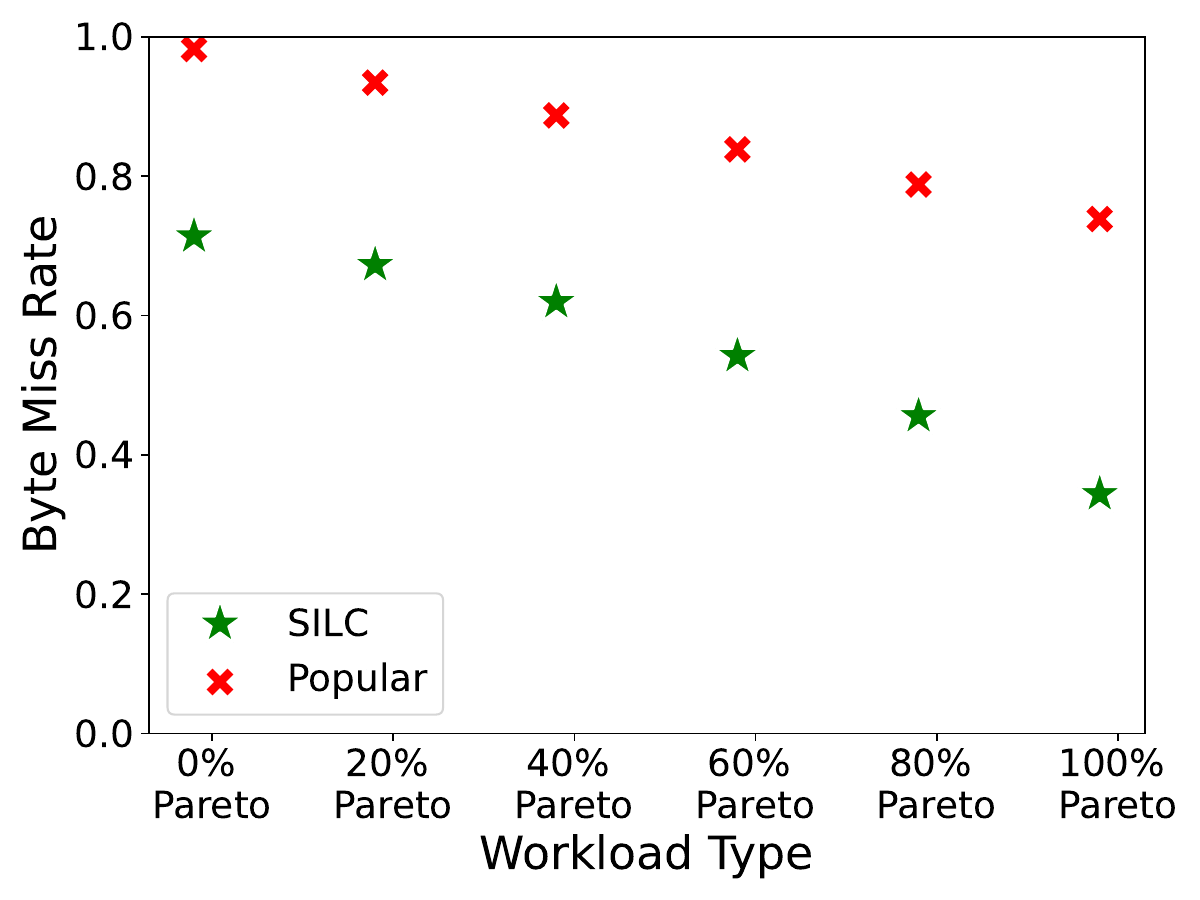}
        \caption{\centering \textit{Vs. Caching Top-K Videos}}
        \label{fig:popbaseline}
\end{subfigure} \hfill
\begin{subfigure}{0.24\linewidth}        
        \includegraphics[width=\linewidth]{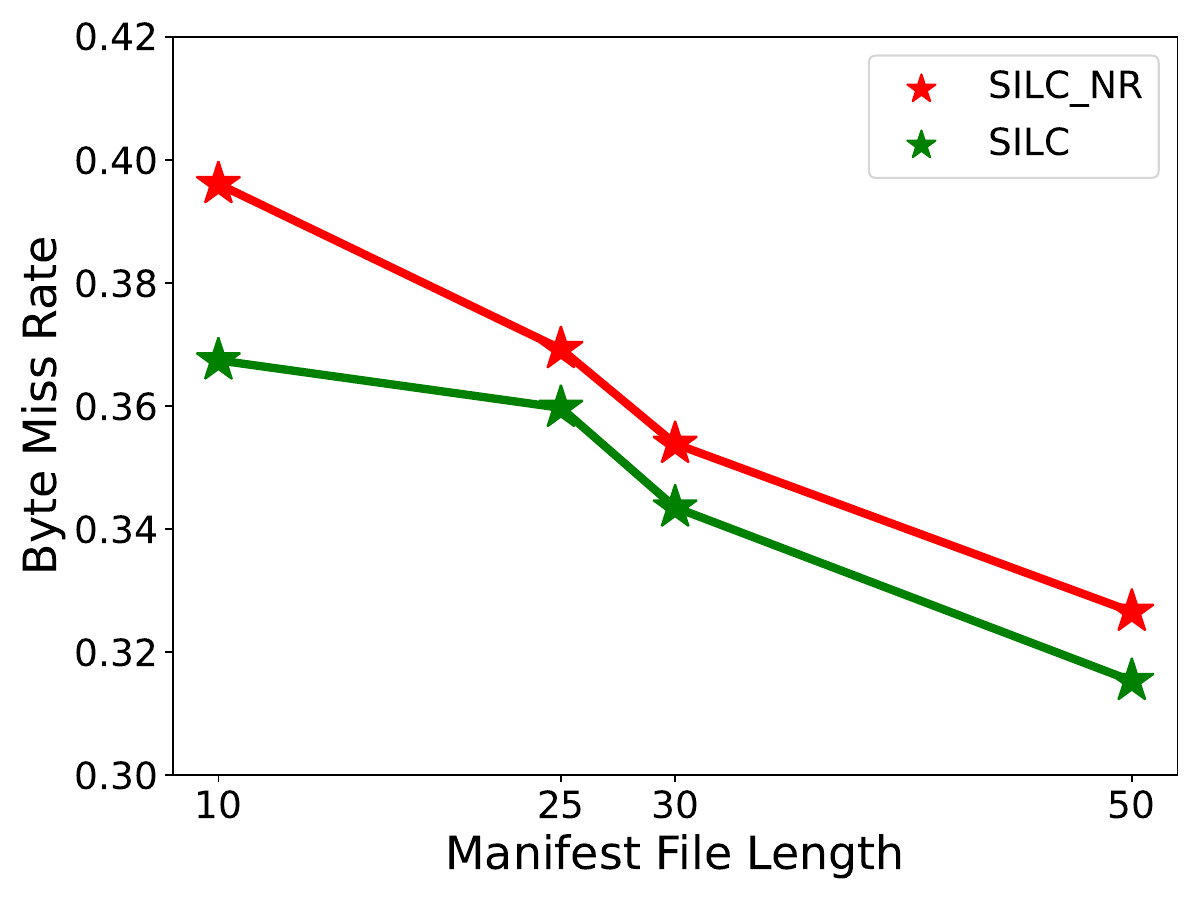}
        \caption{\centering \textit{Length of Manifest Files}}
        \label{fig:mflen}
\end{subfigure} 
    \begin{subfigure}{0.24\linewidth}        
        \includegraphics[width=\linewidth]{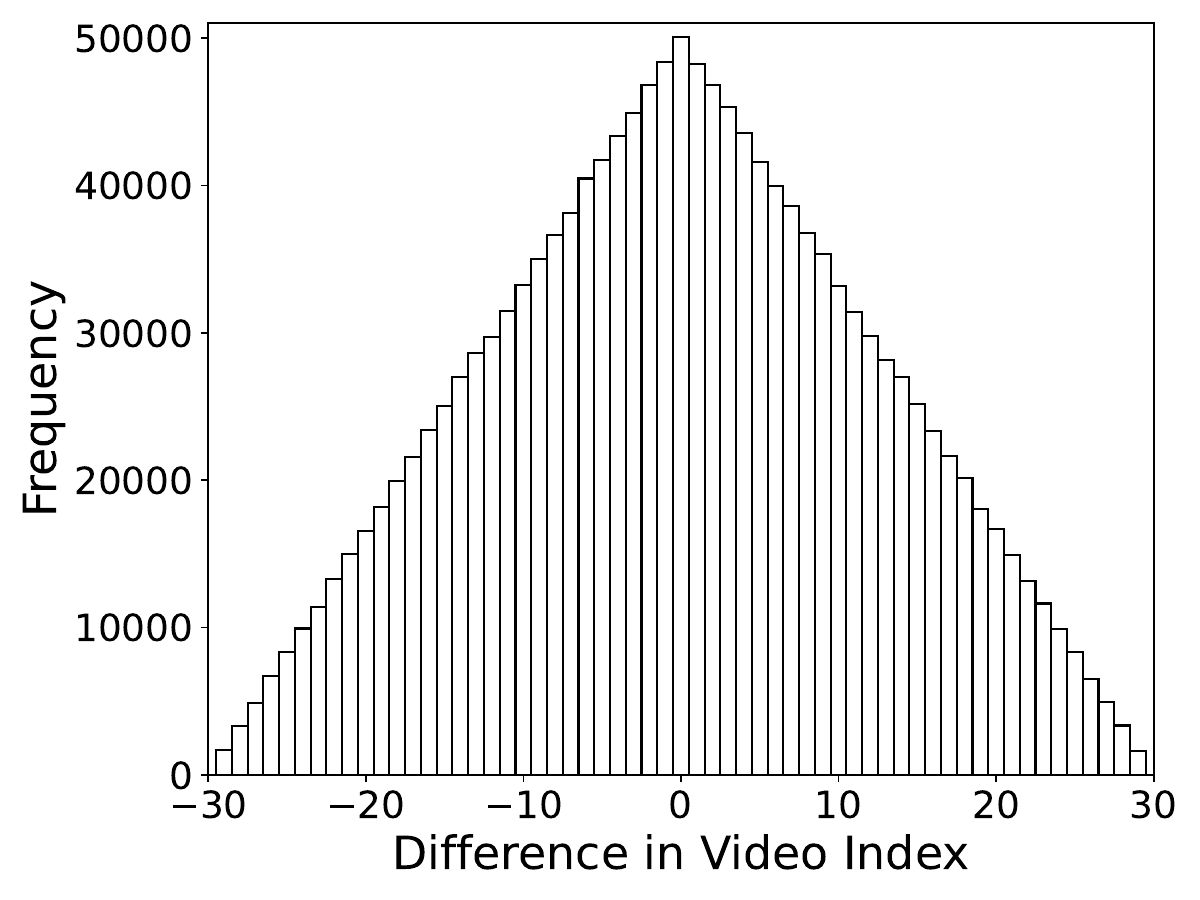}
        \caption{\centering \textit{Reordering Effect}}
        \label{fig:reorderingcount}
\end{subfigure}\hfill

      \vspace{-0.1in}
    \caption{\centering \textbf{Impact of Cache Size, Caching Strategy, Reordering Policy, and Length of Manifest Files on \name.}}\vspace{-0.15in}
\end{figure*}

\para{Varying Cache Size:}
We evaluate the impact of cache size on the performance of \name\ and our baselines. In Fig.~\ref{fig:cache40}, we use the 100\% Pareto sampling work for comparison at 10GB, 50GB, 100GB, 500GB and 1TB of cache size for the entire CDN cluster. We observe that when the cache size is small at 10 GB, there is a need to fetch the majority of requests from the server to the CDN. \name\ achieves 62\% byte miss rate (i.e., need to fetch 62\% traffic using midgress) while the others are above 80\%. At 100 GB, all policies improve as the cache can hold more videos. \name\ has a byte miss ratio of 35\%, LeCaR, LRU and LFUDA close to 40\%, and GDSF achieves 50\%. As the cache size grows to 1 TB, it becomes sufficient to hold most videos and so all caching policies converge.

\para{What if we just cache the most popular videos?} Fig. \ref{fig:popbaseline}
shows the difference in byte miss reduction between a naive implementation of a cache that caches the most popular videos and \name. With a 2.65M dataset, top 20\% videos (following he 80-20 rule in Pareto distributions) account for 80\% of the views. With a 100GB CDN cluster size, we can only store a fraction of this number. This is why the midgress achieved by caching only the popular videos is limited. 
Short video systems undergo significant churn---the set of most popular videos changes over time, and also new users join. 

Thus, even with a larger cache, we do not expect a
top-k popular video caching scheme to perform well since the number of popular videos is huge and quickly changing.

\para{Ablation---Impact of Reordering: }
Fig.~\ref{fig:mflen} compares a \name-variant, called SILC\_NR that \textit{disables} manifest file reordering, with SILC. We observe a robust (but limited) improvement due to the reordering component across various levels of lookahead. Reordering contributes to around 20\% of the gains that \name\ has over the nearest baseline (LeCaR). This implies that the ability to lookahead into the future is the primary driver for \name's gains followed by reordering.

Fig.~\ref{fig:mflen} also plots the byte miss rate for multiple manifest file lengths. The length of a user's manifest file relates to the future knowledge available for \name's cache to make decisions. As the length of the manifest file increases, \name\ performs better overall due to the ability look further ahead into the future. This is true for both \name\ and \name\_NR.

\para{Are some videos reordered more than others?}

Fig.~\ref{fig:reorderingcount} shows how the displacement of a video (actual position minus original position) is  distributed. 
The symmetry of the distribution  implies that content creators and publishers will not see significant variation in their views due to the reordering by \name. That is, a video is as likely to move up the order as it is down the order.

%% file: related.tex
\section{Related Work}

\textbf{Short-Video Streaming:} Existing work  
largely focuses on how to reduce the networking footprint between clients and  server \cite{noncdn, li2023dashlet, vidcdn1,vidcdn2}. Some recent works introduce personalized adaptive video streaming \cite{personalizedadaptive}, adaptive prefetching decisions \cite{winwin} and optimizing decision space for ABR algorithms \cite{tladder}. This is  orthogonal to, and can be combined with, \name. 
 
Many have collected large datasets from short video platforms of Douyin (ByteDance's equivalent of Tiktok in China) and Vine!\cite{10.1145/3479239.3485701,svd,mobileinstant}. None of these works are focused on designing a CDN to improve short video delivery. 

\para{Heuristic Caching Algorithms}
This includes work over multiple decades like 
LRU \cite{lru}, SLRU \cite{slru}, 2Q \cite{2q}, LFU \cite{lfu}, TinyLFU \cite{tinylfu}, FIFO \cite{fifo}, LFUDA \cite{lfuda}, CLOCK \cite{clock}, ARC \cite{arc} and Threshold-LRU \cite{threshold-lru}. These algorithms have low overhead and use a combination of recency, frequency and size to make eviction decisions. 

SIEVE \cite{sieve}, a caching policy tailored to Zipfian distributions, uses a FIFO queue and bit to decide evictions. \name\ can work well on a more skewed distribution like the Pareto distribution and keeps track of the popularity of an object using its future frequency count.

\para{Learning-Based Caching Algorithms} These newer policies 
use ML techniques to predict either the popularity, next request time, etc., 
towards a cache eviction metric. The algorithms learn using historical traces. 
Examples include LRB \cite{lrb}, LeCaR \cite{lecar}, LHD \cite{lhd}, CACHEUS \cite{cacheus}, LFO \cite{lfo}.  Adaptsize \cite{adaptsize} and Darwin \cite{darwin} work with the goal of designing the best admission policies to be combined with an LRU cache. RL-Cache \cite{rl-cache} uses reinforcement learning and combines frequency, recency and size to decide whether to admit an object in cache. \name\ only relies on {\it known} future history, and thus does not need to use any ML techniques, does not propose any admission policies, and does not make estimates based on historical traces.   
Hence we do not need GPUs or extra computational resources for training.

\para{Caching Videos:} Past work has highlighted the importance of caching policies, and emphasized the need to create a cache for short video streaming \cite{chen2019study, 10.1145/3386290.3396937}. Midgress minimization has also been targeted across multiple traffic classes \cite{sundarrajan2020midgress}. However, to the best of our knowledge, none of these papers have focused on short videos.  
For long videos, AViC \cite{avic} outperforms its baselines by predicting  arrival times of future requests for following chunks of the same (current) video. AViC also has an admission policy that predicts if requests are popular or not. AViC has limited applicability in our setting since short videos do not contain too many chunks 
and because videos may not be seen entirely,  all chunks may not need to be fetched. 
Further \name\ relies on only known lookahead and does not need to ``guess'' or estimate. 
HALP \cite{halp} designs CDNs for YouTube videos and uses a learnable scoring function on a small set of proposed objects for eviction. \name\ makes a singular decision on the object to evict based on its future frequency. CACA \cite{10.1145/3343031.3350890} admits videos to cache based on metrics like video category, author and duration and not by request patterns. For \name's cache, the most important metric is frequency for near-future requests. 

\para{Distribution of Workloads:} Past work has 
emphasized that most web object popularity matches the {Zipf distribution}, e.g., peer-to-peer file sharing, Web pages, etc. \cite{workload,breslau1999web, gummadi2003measurement}. 

An analysis of Douyin short video platform  \cite{chen2019study} 
found that the distribution of the most popular videos follows Zipf, but the other videos do not.  
Our study and analysis indicates TikTok videos fit 
a Pareto Distribution.

%% file: conclusion.tex
\vspace{-15pt}
\section{Concluding Discussion}

\noindent\textbf{Implications of our research: }Our work has implications for CDN design and systems research. First, \name\ can be deployed today by CDN operators to improve caching efficiency and lower midgress costs. Second, we hope that other system designers will exploit the \textit{lookahead} capability in short video  system  contexts in future work, to meet other goals beyond CDN midgress and cache hits. Finally, our work is unique in that it includes a mix of user study and systems evaluations---we believe that such evaluations need to  become more commonplace in interactive systems.

\para{Limitations of our research: } First, our work offers a \textit{best-effort} look into TikTok's system architecture. Yet, many pieces of TikTok remain black box, e.g., the recommendation algorithm. For these aspects, we emulate a variety of distributions and evaluated our work across them. Our user study group was also limited to users on our campus and those that use Prolific. While this population has some skew (commonality), we believe that the TikTok populace contains several such groups, and thus our experiments  are representative of overall TikTok behavior and possible benefits from \name.

\para{Reordering videos: }This paper is focused on reducing CDN costs and user-facing aspects---these are the two aspects that we can control, given that the recommender and the content provider are black boxes. Naturally, the recommender may impose ordering among the videos, but the point of our results is that any such order (if it exists) does not appear to matter in terms of user engagement.

\para{Generalization to other short video systems:} While our works focused on TikTok, generalizing \name\ to other short video systems like YouTube Shorts and Instagram Reels may involve subtle design changes (in spite of manifest file equivalents in those systems). 

\para{Ethical Concerns:} Our user study and data collection were both reviewed and approved by our university's Institutional Review Board. We only present aggregate data from TikTok histories and no user-specific data.  

%% file: appendix.tex
\section*{Appendix}

\subsection*{Additional User Study Details} \label{appendix:userstudy}
In Table~\ref{tab:tiktokstudy}, we list the different questions that we asked participants of our user study at the end of a watch session along with the corresponding Likert scales. 

\begin{table*}
\centering
\begin{tabular}{|m{7cm}|m{3.5cm}|}
\hline
\textbf{Question} & \textbf{Scale} \\ \hline
I was more engaged by session \_\_\_\_ & 1-Session 1 \newline\ 7-Session 2 \\ \hline
In session \_\_\_\_, I saw videos that were closer to the kinds of videos I normally saw in my past TikTok viewing. &   1-Session 1 \newline 7-Session 2 \\ \hline
I was more satisfied by session \_\_\_\_ & 1-Session 1 \newline 7-Session 2 \\ \hline
I was more absorbed in session \_\_\_\_ & 1-Session 1 \newline 7-Session 2 \\ \hline
I felt more frustrated in session \_\_\_\_ & 1-Session 1 \newline 7-Session 2 \\ \hline
I found my experience more rewarding in session \_\_\_\_ & 1- Session 1 \newline 7-Session 2 \\ \hline
I felt more interested in session \_\_\_\_ & 1- Session 1 \newline 7-Session 2 \\ \hline
Rate session 1 based on the content quality of the videos that you saw. Note that we are entirely referring to the content of the videos. & 1-Very Dissatisfied \newline 7- Satisfied \\ \hline
Rate session 2 based on the content quality of the videos that you saw. Note that we are entirely referring to the content of the videos. & 1-Very Dissatisfied \newline 7- Satisfied \\ \hline
Rate session 1 based on the network conditions and/or technical issues that you experienced. Note that we are referring to the buffering, lag, network issues, technical concerns etc. only. & 1-Very Poor \newline 7- Excellent \\ \hline
Rate session 2 based on the network conditions and/or technical issues that you experienced. Note that we are referring to the buffering, lag, network issues, technical concerns etc. only. & 1-Very Poor \newline 7- Excellent \\ \hline
How engaging was session 1 for you? & 1-Very Boring \newline 7- Very Engaging \\ \hline
How engaging was session 2 for you? & 1-Very Boring \newline 7- Very Engaging \\ \hline
\end{tabular}
\caption{\textbf{User Study Questions and Responses.}}
\label{tab:tiktokstudy}
\end{table*}

\begin{table}[h!]
\centering
\resizebox{\columnwidth}{!}{%
\begin{tabular}{|>{\centering\arraybackslash}m{2cm}|>{\centering\arraybackslash}m{2cm}|>{\centering\arraybackslash}m{5cm}|}
\toprule
\hline
\hline
 & Heuristic & Description \\ \hline
\midrule
LRU \cite{lru} & Recency & Evicts the least recently used object  \\ \hline
FIFO \cite{fifo} & Recency & Evicts objects based on the order they enter the cache \\ \hline
LFU \cite{lfu} & Frequency & Evicts the least frequent object \\ \hline
LFUDA \cite{lfuda} & Frequency & LFU with dynamic aging to react to shifts in the popular objects across time\\ \hline
GDSF \cite{gdsf} & Frequency and Size & Evicts based on frequency and object size\\ \hline
Random & Random & Randomly removes objects from the cache \\ \hline\hline
\bottomrule
\end{tabular}%
}
\caption{\centering \textbf{Description of the heuristic evictions policies we use as baselines.}}
\label{table:heuristic}
\end{table}

\begin{table}[h!]
\centering
\resizebox{\columnwidth}{!}{ %
\begin{tabular}{|>{\centering\arraybackslash}m{2cm}|>{\centering\arraybackslash}m{3cm}|>{\centering\arraybackslash}m{3cm}|}
\toprule \hline \hline
& Modifications & Description \\ \hline
LeCaR \cite{lecar} & Changing the configurations did not impact the results significantly & Uses regret minimization to combine recency and frequency\\ \hline
LRB \cite{lrb} & We used author provided scripts to tune the algorithm's major hyperparameter (memory window) & Approximate the Belady MIN algorithm and evicts an object that will be requested beyond a threshold \\ \hline
LHD \cite{lhd} & Changing the configurations did not impact results significantly & Predicts the hit density of objects to cache those that contribute most to the hit rate given their size \\ \hline
AdaptSize \cite{adaptsize} & We adjusted the initial threshold size to 6.5 MB based on tuning the algorithm to our traces & Combines admission control with LRU using a dynamic size thresold\\ \hline\hline
\bottomrule
\end{tabular}%
}
\caption{\centering \textbf{Description of the learning-based evictions policies we use as baselines.}}
\label{table:learning}
\end{table}
\subsection*{Description of Baselines}
Information about the heauristic and learning-based baselines we use for evaluation is in Tables \ref{table:heuristic} and \ref{table:learning}, respectively.

\subsection*{Evaluation Results}
Evaluation of \name\ in terms of byte hit and object miss rate is presented in Tables \ref{table:bytehit} and \ref{table:objmiss}, respectively.

\begin{table*}
\renewcommand{\arraystretch}{1.3}
\resizebox{\textwidth}{!}{
\begin{tabular}
{c|>{\columncolor[HTML]{8FBC8F}}c>{\centering\arraybackslash}m{1.7cm}>{\centering\arraybackslash}m{1.7cm}>{\centering\arraybackslash}m{1.7cm}>{\centering\arraybackslash}m{1.7cm}>{\centering\arraybackslash}m{1.7cm}>{\centering\arraybackslash}m{1.7cm}>{\centering\arraybackslash}m{1.7cm}>{\centering\arraybackslash}m{1.7cm}>{\centering\arraybackslash}m{1.7cm}>{\centering\arraybackslash}m{1.7cm}>{\centering\arraybackslash}m{1.7cm}}
\toprule
\hline
\hline
Workload Type & SILC (Ours) & LeCaR (LRU+LFU) & LRB (Approx. Belady) &  FIF \S\ref{sec:simulation} (Practical Belady) & LRU (Recency)&  LFUDA (Frequency)&  FIFO (Recency)&  Random &  GDSF (Frequency+Size)&  LHD (Hit Density)&  AdaptSize (Size) &   LFU (Frequency)\\
\hline
0\% Pareto & 28.66\% & 18.92\% (-51.5\%) & 19.76\% (-45.0\%) & 19.89\% (-44.1\%) & 18.79\% (-52.5\%) & 18.75\% (-52.9\%) & 18.17\% (-57.7\%) & 16.96\% (-69.0\%) & 15.02\% (-90.7\%) & 14.43\% (-98.6\%) & 12.25\% (-133.9\%) & 8.59\% (-233.8\%) \\ \hline
20\% Pareto & 32.75\% & 23.40\% (-40.0\%) & 23.76\% (-37.8\%) & 24.03\% (-36.3\%) & 23.12\% (-41.7\%) & 23.30\% (-40.6\%) & 21.70\% (-50.9\%) & 20.41\% (-60.5\%) & 18.95\% (-72.8\%) & 17.91\% (-82.9\%) & 12.91\% (-153.6\%) & 11.67\% (-180.7\%) \\ \hline
40\% Pareto & 38.07\% & 29.36\% (-29.7\%) & 31.19\% (-22.0\%) & 29.62\% (-28.5\%) & 28.95\% (-31.5\%) & 29.08\% (-30.9\%) & 26.57\% (-43.3\%) & 25.12\% (-51.5\%) & 24.17\% (-57.5\%) & 22.40\% (-69.9\%) & 17.01\% (-123.8\%) & 15.91\% (-139.3\%) \\ \hline
60\% Pareto & 45.83\% & 38.14\% (-20.2\%) & 37.13\% (-23.4\%) & 37.99\% (-20.6\%) & 37.63\% (-21.8\%) & 37.16\% (-23.3\%) & 34.01\% (-34.7\%) & 32.36\% (-41.6\%) & 31.51\% (-45.4\%) & 29.08\% (-57.6\%) & 22.84\% (-100.6\%) & 20.86\% (-119.7\%) \\ \hline
80\% Pareto & 54.45\% & 47.91\% (-13.6\%) & 46.77\% (-16.4\%) & 47.46\% (-14.7\%) & 47.31\% (-15.1\%) & 45.95\% (-18.5\%) & 42.56\% (-27.9\%) & 40.64\% (-34.0\%) & 39.62\% (-37.4\%) & 36.71\% (-48.3\%) & 29.38\% (-85.3\%) & 24.67\% (-120.7\%) \\ \hline
100\% Pareto & 65.65\% & 60.43\% (-8.6\%) & 60.07\% (-9.3\%) & 60.05\% (-9.3\%) & 60.01\% (-9.4\%) & 56.63\% (-15.9\%) & 54.68\% (-20.1\%) & 52.07\% (-26.1\%) & 49.66\% (-32.2\%) & 47.49\% (-38.2\%) & 37.41\% (-75.5\%) & 27.44\% (-139.3\%) \\\hline\hline
\bottomrule 
\end{tabular}}
\caption{\centering \textbf{Byte Hit Rate.} \textit{Byte hit rate comparison for all workloads and baselines. Values in brackets indicate percentage difference with respect to \name.}}
\label{table:bytehit}
\end{table*}

\begin{table*}
\renewcommand{\arraystretch}{1.3}
\resizebox{\textwidth}{!}{
\begin{tabular}
{c|>{\columncolor[HTML]{8FBC8F}}c>{\centering\arraybackslash}m{1.7cm}>{\centering\arraybackslash}m{1.7cm}>{\centering\arraybackslash}m{1.7cm}>{\centering\arraybackslash}m{1.7cm}>{\centering\arraybackslash}m{1.7cm}>{\centering\arraybackslash}m{1.7cm}>{\centering\arraybackslash}m{1.7cm}>{\centering\arraybackslash}m{1.7cm}>{\centering\arraybackslash}m{1.7cm}>{\centering\arraybackslash}m{1.7cm}>{\centering\arraybackslash}m{1.7cm}}
\toprule
\hline
\hline
Workload Type & SILC (Ours) & LeCaR (LRU+LFU) & LRB (Approx. Belady) &  FIF \S\ref{sec:simulation} (Practical Belady) & LRU (Recency)&  LFUDA (Frequency)&  FIFO (Recency)&  Random &  GDSF (Frequency+Size)&  LHD (Hit Density)&  AdaptSize (Size) &   LFU (Frequency)\\
\hline
0\% Pareto & 71.15\% & 80.89\% (13.7\%) & 80.06\% (12.5\%) & 79.94\% (12.4\%) & 82.10\% (15.4\%) & 81.04\% (13.9\%) & 81.68\% (14.8\%) & 82.86\% (16.4\%) & 69.62\% (-2.2\%) & 68.78\% (-3.3\%) & 73.34\% (3.1\%) & 91.26\% (28.3\%) \\\hline
20\% Pareto & 66.30\% & 75.63\% (14.1\%) & 75.37\% (13.7\%) & 75.03\% (13.2\%) & 75.80\% (14.3\%) & 75.57\% (14.0\%) & 77.49\% (16.9\%) & 78.75\% (18.8\%) & 64.83\% (-2.2\%) & 64.18\% (-3.2\%) & 72.70\% (9.6\%) & 87.24\% (31.6\%) \\\hline
40\% Pareto & 60.47\% & 69.11\% (14.3\%) & 67.65\% (11.9\%) & 68.91\% (13.9\%) & 69.56\% (15.0\%) & 69.18\% (14.4\%) & 72.16\% (19.3\%) & 73.55\% (21.6\%) & 59.30\% (-1.9\%) & 58.82\% (-2.7\%) & 66.59\% (10.1\%) & 82.39\% (36.2\%) \\\hline
60\% Pareto & 52.14\% & 59.69\% (14.5\%) & 61.28\% (17.5\%) & 59.88\% (14.8\%) & 60.23\% (15.5\%) & 60.45\% (15.9\%) & 64.11\% (22.9\%) & 65.71\% (26.0\%) & 51.75\% (-0.8\%) & 51.33\% (-1.6\%) & 58.77\% (12.7\%) & 76.67\% (47.0\%) \\\hline
80\% Pareto & 43.62\% & 49.99\% (14.6\%) & 51.79\% (18.7\%) & 50.46\% (15.7\%) & 50.61\% (16.0\%) & 51.77\% (18.7\%) & 55.55\% (27.3\%) & 57.37\% (31.5\%) & 43.99\% (0.8\%) & 43.43\% (-0.5\%) & 50.56\% (15.9\%) & 72.71\% (66.7\%) \\\hline
100\% Pareto & 32.62\% & 37.64\% (15.4\%) & 38.51\% (18.0\%) & 38.03\% (16.6\%) & 38.68\% (18.6\%) & 41.14\% (26.1\%) & 43.50\% (33.3\%) & 45.97\% (40.9\%) & 34.29\% (5.1\%) & 33.12\% (1.5\%) & 40.79\% (25.0\%) & 69.68\% (113.6\%) \\ \hline\hline
\bottomrule
\end{tabular}}
\caption{\centering \textbf{Object Miss Rate.} \textit{Object miss rate comparison for all workloads and baselines. Values in brackets indicate percentage difference with respect to \name.}}
\label{table:objmiss}
\vspace{-20pt}
\end{table*}